\documentclass{SCIS2020cn}

\usepackage{siunitx}

\begin{document}
\ArticleType{评述}
\newcommand{\teff}{\mbox{$T_{\rm eff}$}} 
\newcommand{\logg}{{\rm{log}~$g$}}
\newcommand{\feh}{{\rm [Fe/H]}} 
\Year{2020}
\Vol{50}
\No{1}
\BeginPage{1}
\DOI{}
\ReceiveDate{}
\ReviseDate{}
\AcceptDate{}
\OnlineDate{}

\title{大视场测光巡天流量定标方法}{}

\entitle{Photometric calibration methods for wide-field photometric surveys}{}

\author[1]{黄博闻}{}
\author[1]{肖凯}{}
\author[1]{苑海波}{yuanhb@bnu.edu.cn}

\enauthor[1]{Bowen HUANG}{}
\enauthor[1]{Kai XIAO}{}
\enauthor[1]{Haibo YUAN}{yuanhb@bnu.edu.cn}

\address[1]{北京师范大学天文系，100875，北京}

\enaddress[1]{Department of Astronomy, Beijing Normal University, Beijing {\rm 100875}, China}

\Foundation{基金资助：国家自然科学基金No. 12173007，No. 11603002；科技部重大基础研发计划No. 2019YFA0405503；中国载人航天工程巡天空间望远镜专项科学研究NO. CMS-CSST-2021-A08 和 NO. CMS-CSST-2021-A09}

\AuthorMark{黄博闻, 肖凯, 苑海波}

\AuthorCitation{黄博闻, 肖凯, 苑海波}
\enAuthorCitation{Huang Bowen, Xiao Kai, Yuan Haibo}

\abstract{均匀且精确的流量定标是大视场测光巡天的难点和成功的关键因素之一。
本文总结了当前大视场巡天流量定标方法的发展情况，包括经典的标准星方法，
“硬件驱动/观测驱动”的Ubercalibration、Hypercalibration、Forward Global Calibration方法，
及“软件驱动/物理驱动”的Stellar Locus Regression、Stellar Locus及Stellar Color Regression方法，并对这些方法的优势、限制及其后续发展, 特别是Stellar Color Regression方法，进行评述与讨论。}

\enabstract{
Photometric calibration methods for wide-field photometric surveys
Uniform and accurate photometric calibration plays an important role in the current and next-generation wide-field imaging surveys. Herein, we review the modern photometric calibration methods, including the classic standard star method, “hardware/observation-driven” methods (such as the Ubercalibration, Hypercalibration, and Forward Global Calibration Methods), and “software/physics-driven” methods (e.g., the Stellar Locus Regression, Stellar Locus, and Stellar Color Regression Methods). Further, we discuss their advantages, limitations, and future developments toward millimagnitude precision calibration.}


\enkeywords{photometric surveys, flux calibration, stellar parameters, interstellar extinction}

\maketitle

\section{引言} 

天文学是以观测为驱动、以发现为导向的实测学科。随着望远镜、探测器
、多目标光谱仪和数据处理技术的进步，以SDSS\cite{York et al.(2000)}计划为代表的大视场测光与光谱巡天在天文学的研究中发挥着越来越重要的作用。近二十年来，大视场大规模巡天计划层出不穷，如国内的
Large Sky Area Multi-Object Fiber Spectroscopic Telescope（LAMOST）\cite{Luo et al.(2015)} \cite{Deng et al.(2012)}\cite{Liu et al.(2014)}\cite{Zhao et al.(2012)}、
Beijing-Arizona-Taipei-Connecticut（BATC）Sky Survey\cite{Fan et al.(1996)}、
South Galactic Cap of u-band Sky Survey（SCUSS）\cite{Zhou et al. SCUSS1}\cite{Zou et al. SCUSS2}\cite{Zou et al. SCUSS3}、
Xuyi Schmidt Telescope Photometric Survey of the Galactic Anti-center （XSTPS-GAC）\cite{Liu et al.(2014)}、
Beijing-Arizona Sky Survey（BASS）\cite{Zou et al.(2017)} 、
Stellar Abundance and Galactic Evolution（SAGE）\cite{Zheng et al.(2018)}\cite{Zheng et al.(2019)}
和国际上的
Sloan Digital Sky Survey（SDSS）\cite{York et al.(2000)} I-V、
Galaxy Evolution Explorer（GALEX）\cite{Martin et al.(2005)}、
Two Micron All Sky Survey（2MASS）\cite{Skrutskie et al.(2006)}、
Wide-field Infrared Survey Explorer（WISE）\cite{Wright et al.(2010)}、
Dark Energy Survey（DES）\cite{DES(2005)}、
Pan-STARRS1 Survey（PS1）\cite{Schlafly(2012)}、
Skymapper\cite{Keller et al.(2007)}和
{\it Gaia}\cite{Gaia(2016)}等。
这些已完成或接近完成的巡天计划在恒星物理、银河系结构与近场宇宙学、星系形成与演化、宇宙大尺度结构、暗物质与暗能量、时域天文、稀有天体或现象的发现等天文学重大领域发挥了极其重要的作用。而即将开展或计划中的下一代国内外测光与光谱巡天计划，如
Vera Rubin Observatory（LSST）\cite{LSST(2019)}、
Javalambre-Physics of the Accelerated Universe Astrophysical Survey（JPAS）\cite{Benitez et al.(2014)}、
Nancy Grace Roman Space Telescope（WFIRST）\cite{Green et al.(2012)}、
EUCLID\cite{Laureijs et al.(2011)}、
Dark Energy Spectroscopic Instrument（DESI）\cite{DESI Collaboration et al.(2016)}、
WEAVE\cite{Dalton et al.(2012)}、
Maunakea Spectroscopic Explorer（MSE）\cite{McConnachie et al.(2016)}等，
以及国内的
China Space Station Telescope（CSST）\cite{Zhan(2021)}、
Wide Field Survey Telescope（WFST）\cite{Lou et al.(2016)}、
Multi-channel Photometric Survey Telescope（Mephisto）\cite{Mephisto(2020)}、
LAMOST二期工程、
司天工程 \cite{Liu et al.(2021)}
等，更将对天文学的发展产生重大影响。

对大视场测光巡天来讲，保证望远镜在不同条件下，在探测器不同位置上，在不同时刻观测下的相距很远的目标之间 流量测量的一致性，即均匀的流量定标，是影响巡天成功的最关键因素之一。流量定标精度与巡天面积、巡天深度、空间分辨本领、滤光片、时间采样等因素一起决定了其巡天发现空间的大小。
流量定标精度通常决定了天体流量测量误差的下限，对Ia型超新星宇宙学、星系大尺度成团性研究、
恒星测光参数测定、星系测光红移测量、及特殊天体（如类星体、极端贫金属星）的筛选等具有重要影响\cite{Kent et al.(2009)}。

流量定标也是大视场巡天数据处理的难点所在。由于以下三种主要因素的存在，地面大视场测光巡天流量定标存在一个“1\%精度的瓶颈”（1\% precision barrier\cite{Stubbs and Tonry(2006)}）。一是由于理想的测光夜并不存在，地球大气的不透明度存在短时标（秒到分钟的量级）的快速变化，无法用单个不变或缓变的大气消光系数精确改正大气不透明度的影响，
且不同类型恒星受到大气的影响也不尽相同；二是由于仪器的系统误差很难改正，特别是平场改正导致的误差。理想的平场改正需要一个光路及能谱分布均与观测目标源一致的均匀面光源，而不管是圆顶平场、天光平场还是夜天光平场都有其固有的缺陷；三是
在拼接相机多通道读出的情况下由于探测器电子学的不稳定性所导致的（如增益）变化。

本文结构如下：本章介绍流量定标的概念、经典的标准星方法及其局限、地面与空间流量定标的比较；
第2章介绍并评述“硬件驱动/观测驱动”的大视场测光
巡天定标方法，包括
Ubercalibration\cite{Padmanabhan(2008)}、Hypercalibration\cite{Finkbeiner(2016)}以及
Forward Global Calibration Method（FGCM）\cite{Burke(2018)}方法；
第3章介绍并评述“软件驱动/物理驱动”的大视场测光
巡天定标方法，包括
Stellar Locus Regression (SLR\cite{High et al.(2009)}）、
Stellar Locus (SL\cite{Lopez-Sanjuan et al.(2019)}）以及 
Stellar Color Regression (SCR\cite{Yuan et al.(2015a)})方法；
第4章介绍SCR方法在大视场巡天定标中的应用情况并对
其未来的发展进行讨论；最后进行总结与展望。

\subsection{什么是流量定标}
\subsubsection{星等}
星等最早由古希腊天文学家喜帕恰斯定义，其约定肉眼可见最暗的星为六等星，最亮的星为一等星，中间均匀划分。但是由于人眼的非线性效应，使得最亮的星比最暗的星亮100倍，因此星等值和亮度不是线性关系。在现代天文学中，诺曼·罗伯特·普森首先沿用了这一非线性定义，并将其数学化。星等值越小，亮度越大，且每减小5等亮度增大100倍。然而，这一定义没有很好地解释“亮度”这一概念，其次星等和亮度是对数关系，且仅能反应相对值，如果要想定义天体亮度的绝对值，则需要引入新的定义。在现代天文学中有了对于星等更加标准的定义，常见如AB星等系统，星等可以根据流量得出\cite{Oke(1974)}：
\begin{equation}
    AB = -2.5logF_{\nu}-48.60
    \label{Equ:Standardsystem1}
\end{equation}
其中，$F_{\nu}$是天体的流量密度，单位是$ergs \, cm^{-2} \, s^{-1} \, Hz^{-1}$。

将上述定义扩展至宽波段测光后，AB星等可由下式定义
\cite {Fukugita et al.(1996)}：
\begin{equation}
    m^{obs} = -2.5log_{10} (\frac{\int_{0}^{\infty} F_{\nu}(\lambda) \times S^{obs}(\lambda) \times d \lambda}{\int_{0}^{\infty} S^{obs}(\lambda) \times d \lambda})-48.60
    \label{Equ:Standardsystem3}
\end{equation}
其中，$S^{obs}$是观测通带的透过率，其包含了从地球大气、光学系统到仪器本身的响应。

\subsubsection{测光系统和流量定标}

根据公式\ref{Equ:Standardsystem1}，如果$S^{obs}$可知，那么测光系统可以根据该式精确定义。
但是，因为$S^{obs}$会随着观测条件变化，这种严格的定义是非常困难的。
因此，测光系统通常会由一系列被选定的标准星的星等定义\cite{Sterken and Manfroid(1992)}。根据滤光片带宽不同，测光系统可以分为宽带测光（$\Delta \lambda <   100 nm$）、中带测光（$  7 nm < \Delta \lambda <   40 nm$）和窄带测光（$\Delta \lambda <   7 nm$）\cite{Bessell(2005)}。
因此，标准测光系统需要给定一组标准星以定义其零点。这些标准星通常有多次精确测量，不存在亮度变化，
有宽广的亮度以及颜色范围，且尽量均匀覆盖全部的天空位置。最早的标准测光系统是UBV测光系统 \cite{Johnson and Morgan(1953)}，其标准星包括一些近邻的亮星和疏散星团。

而流量定标便是指，在给定的标准测光系统下，通过改正地球大气吸收以及仪器效应\cite {Bessell(2005)}的影响，将仪器星等转化为大气外的视星等，使其能够反映观测目标的真实亮度，
且能够同其他数据进行比较。对于一幅图像来讲，该过程可以由公式\ref{Equ:Standardsystem2}描述：
\begin{equation}
    m = m_{ADU} + ZPT(t, X, Y, m, color) 
    \label{Equ:Standardsystem2}
\end{equation}
其中，$m$为定标后星等，$m_{ADU}$为仪器星等，$ZPT$为广义的零点。零点可以是时间的函数（如反映大气不透明度/仪器效率的变化），也可能跟源在探测器上的位置（X，Y）有关（如反映平场改正的影响或测光时导致的一些系统效应），
也可能跟源的亮度（如探测器的非线性效应）及颜色（如大气消光的颜色效应）有关，甚至这些变量之间还可以有耦合
（如大气消光颜色效应会导致零点和大气质量与源的颜色均有关系；CCD探测器的Charge Transfer Inefficiency（CTI）效应会导致零点与源的亮度和
在探测器的位置均有关系，滤光片透过曲线随入射角的变化导致的零点与源的颜色和在探测器的位置均有关系）。
在理想情况下（不考虑仪器效应改正误差及大气消光颜色效应），$ZPT$仅为时间的函数，对于一幅给定图像则为一个常数。

需要注意的是，流量定标分为相对定标（内部定标）和绝对定标（外部定标）。
相对定标指的是得到天体的相对流量，使得不同天体的亮度可以在同一个系统内进行比较。
而绝对定标是指得到天体的物理流量，此时需要根据系统定义的通带精确确定测光系统的零点，
使得定标后星等够反映准确的物理流量。
由于天体的精确距离难以测量，大部分情况下人们更关心天体的相对流量大小\footnote{ 在需要知道天体绝对流量以测量其距离时（如Ia型超新星），
或已知其（宇宙学）距离需要测量其光度（如星系）时，
或需要观测与理论模型进行比较时等情况下，绝对定标非常重要。}。
由于天体的大部分内禀性质
（如恒星的基本参数和星系的测光红移）取决于其能谱分布（SED），相比于天体的流量，人们绝大部分情况下更关心天体颜色。 在没有准确的绝对流量定标时，尽管测量的颜色不能反映天体真实的颜色，但这种系统偏差并不会影响天体大部分内禀物理性质的测量。
因此，本文中流量定标除非特别说明均指的是相对定标。

\subsection{标准星方法}
在天文观测中会经常使用标准星进行（绝对）流量定标，如UBVRI测光系统标准星\cite{Landolt(1992)} \cite{Landolt(2009)} \cite{Landolt(2013)} \cite{Stetson(2000)} \cite{Clem & Landolt(2013)}。这些标准星的亮度已知并拥有良好的内部一致性，因此当观测这些标准星时，就可以通过对比当前以及标准星原本测光结果的差异来测量大气吸收系数和仪器零点。在测光夜的情况下，测得的大气吸收系数和仪器零点可以应用到其它天区的定标当中。

标准星方法会考虑仪器零点和简单的大气消光模型，其通常包含如下步骤：
\begin{enumerate}
    \item 
     对仪器星等进行第一步改正，以改正曝光时间和大气吸收的影响，其过程如公式\ref{Equ:Standard1}所示
    \begin{equation}
    m' = m_i + 2.5log_{10}(t_{exp}) - k'X 
    \label{Equ:Standard1}
    \end{equation}
    其中$m'$为第一步修正后的仪器星等；$t_{exp}$为曝光时间；$X$为有效大气质量；$k'$为大气消光系数，
    在简单的大气消光模型中仅考虑其一阶项。$k'$可以通过Bouguer方法\cite{Sterken and Manfroid(1992)}，
    即观测在不同时刻不同大气质量下的同一（标准）星并进行最小二乘线性回归得到，
    也可以通过观测（相邻）时刻不同大气质量下的不同标准星得到。

    \item 改正零点及实际测光系统和标准测光系之间的颜色项，其过程如公式\ref{Equ:Standard2}所示。
    \begin{equation}
    M = m' + fc + ZP 
    \label{Equ:Standard2}
    \end{equation}
     其中$M$为最终定标后的星等；$m'$为第一步修正后的仪器星等；$f$为测光系统颜色项，其表征了实际测光系统和标准测光系统之间的偏差，如果两个系统完全一致，则$f$应当为零；$c$为色指数；$ZP$为该标准测光系统的零点。$f$和$ZP$这两个系数
     需要通过观测不同颜色的标准星得到。
\end{enumerate}

国内的BATC巡天便使用了标准星方法进行定标\cite{Yan et al.(2000)}\cite{Zhou et al.(2001)}。BATC巡天利用兴隆60/90cm施密特望远镜及15个特制滤光片对河内和河外目标进行中波段测光。
在定标时，通过多次观测Oke \& Gunn (1983)\cite{Oke and Gunn(1983)}中的4颗标准星，根据仪器星等和标准星原本星等之间的差异，每晚确定一个一阶大气消光系数与仪器零点。 与上述介绍不同的是，其早期定标还包含一个与波段无关但随时间分段线性变化的零点修正项
来改正大气消光变化的影响\cite{Yan et al.(2000)}，
后来替换为随时间变化的大气消光系数项来更好地改正大气消光变化的影响\cite{Zhou et al.(2001)}。

\subsection{标准星方法的局限性}
标准星方法提供了一套完整的地面巡天定标方案，其不但适用于常规天文观测，也可以应用于巡天当中。
但是，由于大气及仪器效应改正的复杂性（标准星数量远远不足），
通过标准星方法开展高精度流量定标也面临很多的困难与挑战：

\begin{figure*}[ht!]
   \centering
   \includegraphics[width=15.cm]{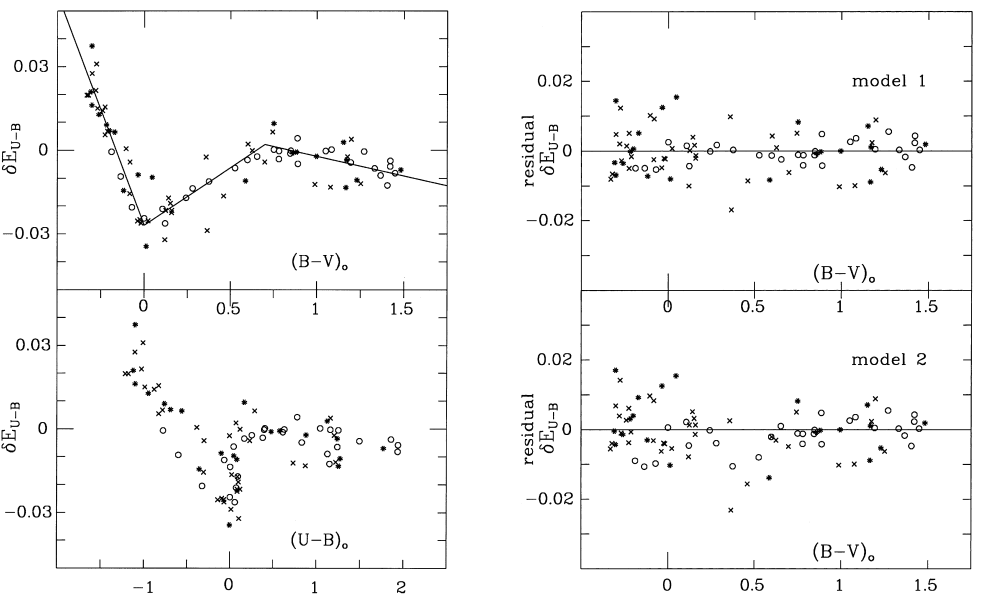}
   \cnenfigcaption{ {\small Balmer Dip对大气消光的影响\cite{Cousins and Caldwell(2001)}。左边两幅图中展示了大气消光同恒星颜色$B-V$和$U-B$的关系，右边两幅图显示了修正后的结果。}} {{\small The effect of Balmer Dip on atmospheric extinction\cite{Cousins and Caldwell(2001)}. The two left panels show the atmospheric extinction as a function of the stellar colors $B-V$ and $U-B$, while the two right panels show the corrected results.}}
  \label{Fig:Cousins2001}
\end{figure*}

\begin{enumerate}

    \item 理想的测光夜是不存在的。地球大气并非一个均匀介质，其包含水蒸气，臭氧以及气溶胶等介质。水汽和臭氧会对紫外和红外波段产生强烈吸收，其分子吸收线非常复杂。对于其余分子和气溶胶来说，其会对光进行散射。
    而且，这些介质的分布存在短时标（秒到分钟的量级）的快速变化。标准星方法由于数量和空间分布的原因很难实时考虑大气消光快速变化所带来的影响。
    
    \item 大气消光的颜色效应。宽带测光的大气消光系数还与恒星颜色有关，需要考虑二阶项的影响。
    由于在特定波段处如巴尔末跳变附近局部SED形状与整体SED形状之间的差异，二阶项形式可能很复杂。 如图\ref{Fig:Cousins2001}所示，Balmer Dip
    在U波段所造成的影响比较复杂，并不能通过一个简单的线性修正加以确定。颜色项也导致精确的大气消光改正需要知道
    天体的（实时）颜色。
    
    \item 大气消光的Forbes效应\cite{Forbes(1842)}。 Forbes效应由J.D.Forbes于1842年发现。 在宽带测光中，大气消光随大气质量的变化并不能用一个线性系数很好地描述，
    其会存在或多或少的弯曲。这种效应在不同波段均会产生，在可见光波段，Forbes效应会导致大气消光系数随着大气质量的增加而逐渐减小\cite{King(1952)}。这是由于随着大气质量的增加，恒星流量的衰减在消光更大的波长处更严重。在近红外波段，Forbes效应非常显著\cite{Manduca and Bell(1979)}，相较于可见光波段其也会变得会非常复杂。这是由于近红外波段处，大气吸收中存在大量且不规则分布的、饱和且重叠的分子线（如水汽）\cite{Young et al.(1994)}。
    
    \item 在实际观测中，需要高精度平场来改正系统响应的不均匀性，特别是大尺度的不均匀性。传统平场改正（圆顶平场、天光平场及夜天光平场）有着各种各样的问题，精度有限。 理想的平场改正需要一个光路及能谱分布均与观测目标源一致的均匀面光源。
    这三种平场都不是完全均匀的，存在亮度梯度\cite{Chromey and Hasselbacher(1996)}。且圆顶平场的能谱分布和光路与观测目标有明显不同；天光平场的拍摄时间有限，
    并依赖于天气情况；夜天光平场的能谱分布与目标源最为接近，但是构建夜天光平场需要长时间的曝光才能获得足够的信噪比，而且需要对不同的位置曝光以扣除图像中的天体信号。
    
     \item 标准星本身的精度也会影响使用标准星定标的精度。标准星本身的精度体现在两个方面，一个是测光的精度，另一个是零点的一致性。
     以Landolt标准星为例\cite{Landolt(1992)}，其包含分布在天赤道上的526颗11.5 < V < 16的UBVRI波段标准星，采用光电倍增管观测多次，测光精度从毫星等到十几个毫星等不等。
     Clem等人后续用CCD探测器对Landolt标准星表进行了扩充\cite{Clem & Landolt(2013)}，将标准星数量扩充至约45000颗，且分布于天赤道与赤纬$-$50度处，其测光精度好于30mmag。
     对于零点一致性来说，我们近期基于后文中提到的恒星颜色回归方法结合{\it Gaia} EDR3 \cite{Gaia(2021)}和LAMOST DR7数据对Landolt 扩展标准星表进行了检验，发现大部分天区零点的一致性在1\%左右，
     但少量天区零点偏差达到0.02星等\footnote{Huang Bowen and Yuan Haibo, in preparation}。
     
    \item 在实际观测中，为实现高精度定标，需要花费大量观测时间拍摄标准星来测量大气消光和仪器零点的变化。

\end{enumerate}

近年来，针对大视场测光巡天流量定标工作，天文学家开发出了一系列新方法。
这些方法可以大致分为两类：一类是“硬件驱动”，或者说“观测驱动”；另一类是“软件驱动”，或者说“物理驱动”。
通过这些方法，我们有机会突破地面大视场巡天百分之一定标精度的限制。
值得注意的是，这些新方法通常更关注的是相对定标（内部定标），其绝对零点仍然需要标准星的观测（外部定标）来确定。

\subsection{空间与地面大视场测光巡天流量定标比较}
与地面大视场测光巡天相比，空间大视场测光巡天流量定标优劣各半。其主要优势体现在不需要考虑复杂的地球大气消光的影响，仪器性能虽然会有长期的变化，但短期内性能稳定。其劣势主要体现在以下几个方面。
一是更复杂的探测器效应改正。由于空间高能辐射/粒子的破坏作用，CCD探测器的CTI效应可能会非常显著，
对数据处理与定标带来了额外的困难\footnote{ 近红外大视场测光巡天的探测器通常不是CCD , CTI效应的影响并不显著（如HST/WFC3, EUCLID/NISP)。}。
另一方面，探测器入轨后性能的变化无法再在实验室进行详细研究，因此要求
对探测器的性能在上天前进行充分的研究，以应对由于某些未知因素导致的性能变化。
二是平场拍摄更加困难。常规平场虽然能对地球或利用平场灯进行拍摄，但空间不均匀性较差，光路亦有差异。
夜天光平场由于像素分辨率高与天光背景暗（特别是在只能空间观测的紫外波段）而难以获取，且也存在一定的梯度。因此，大尺度响应的
不均匀性更加依赖于恒星平场来进行改正。
三是由于通常情况下空间分辨率高的原因\footnote{ 部分空间测光巡天（如GALEX、WISE）的空间分辨本领是相对偏低的。}，单个探测器视场相比地面偏小，（定标）源的数目偏小。 但相应的是，空间望远镜的观测策略也常常会进行针对性的优化，比如Gaia采用多次重复扫描模式，
EUCLID采用4次抖动观测模式（dither pattern）,  CSST采用两个或四个“相同”滤光片重复观测及同一滤光片相邻天区之间存在少许重叠模式。在这些观测模式下，
下节即将介绍的基于重叠区域的Ubercalibration技术将提高流量定标的精度。
四是由于条件变化，测光系统可能与地面实测的有所不同，可能会随时间变化，
绝对流量定标需要对在轨系统响应曲线精确测量\cite{Weiler(2018)}\cite{Maiz Apellaniz and Weiler(2018)}。
值得强调的是，不论是标准星方法，还是下文提及的新方法，虽然是针对更为常见的地面巡天所开发，
但其原理均适用于空间大视场巡天流量定标\cite{Markovc et al.(2017)}\cite{Davini et al.(2021)}\cite{Niu et al.(2021a)}。

\section{硬件驱动的方法}
其中“硬件驱动”的方法基于对大视场成像观测的进一步认识，其包括
Ubercalibration\cite{Padmanabhan(2008)}、Hypercalibration\cite{Finkbeiner(2016)}以及
Forward Global Calibration Method（FGCM）\cite{Burke(2018)}。本节将分别介绍这三种方法。

\subsection{Ubercalibration方法} 
Ubercalibration方法最早为进一步提升 SDSS 测光巡天流量定标精度所开发，取得了巨大的成功，后来被多个大视场巡天广泛使用，包括PS1巡天 \cite{Schlafly(2012)}，盱眙施密特望远镜反银心方向测光巡天\cite{Liu et al.(2014)}，BASS巡天 \cite{Zou et al.(2017)}\cite{ZhouZhimin(2018)}，和{\it Gaia}巡天\cite{Gaia(2016)}等。该方法的基本假设非常简单：相同源在不同观测条件下测得的物理星等应该相同。
这个假设对于绝大多数源都是成立的。
给定一个有足够多重复观测源的巡天数据，在针对性建立该巡天定标模型并对模型参数化的基础上，基于上述简单假设即可对定标参数进行限制和优化，对于不同的巡天项目来说，其定标参数各有不同。

\subsubsection{Ubercalibration方法在SDSS巡天中的应用} 
Ubercalibration方法在SDSS巡天中的定标模型由下式表示：

\begin{equation}
    ZPT = a(t) - k(t)x + f(i,j;t) + ...
\end{equation}

在上式中，Ubercalibration方法考虑了一个相对简单的唯像模型以估计广义零点$ZPT$。该模型类似于泰勒展开的一阶项，没有考虑其它更为细致的因素。其中，$a$项代表光学系统和探测器的响应；$k$项代表地球大气消光系数，$x$为大气质量；$f$项代表平场，其中$i,j$为CCD坐标下的位置；$t$为时间。在上述模型中，$ZPT$的每一部分相互独立且可能随时间变化。
Ubercalibration方法通过重复观测来限制并确定这些参数。

\begin{figure}
    \centering
   \includegraphics[width=8cm]{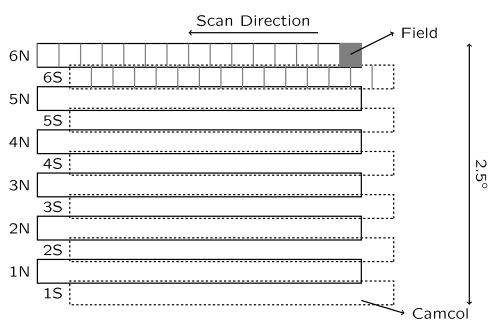}
   \cnenfigcaption{{\small SDSS巡天的观测模式\cite{Padmanabhan(2008)}。SDSS的相机有6列CCD，每一列排布着5种滤光片。在观测时，其沿大圆进行漂移扫描。© AAS. Reproduced with permission.}} {{\small  Geometry of the SDSS imaging\cite{Padmanabhan(2008)}. The SDSS camera has six rows of CCDs with five filters lined up in each row. During the observation, it drift scans along the great circle.
   © AAS. Reproduced with permission.}}
  \label{Fig:SDSS}
\end{figure}

根据如图\ref{Fig:SDSS}所示的SDSS巡天扫描观测模式，
Ubercalibration方法对$ZPT$的每一项进行了进一步简化处理。
$a$项对于每一块CCD均是独立的，$k$项仅与时间有关，$f$项的二维平场退化为一维。
考虑到光学系统和探测器的响应变化较为缓慢，可以假设$a$项在同一个测光夜之内是不变的。
考虑到平场只有在相机维护时才会发生较大且不连续的变化，可以认为$f$项在每次相机维护之间（平场季）是不变的，
每个平场季的时长从一个月到一年不等。地球大气的消光变化相对复杂。这里仅考虑通常测光夜情况下，
每晚大气的消光系数会随时间逐渐变小，其速率约为1 mmag$hr^{-1}$，
因此假设每晚上大气消光系数随时间线性变化，不同波段的变化速率比较接近，同一波段在不同晚上
速率相同。

综上，即可遵从公式\ref{Equ:Uber2}依据重复观测源的亮度相同对参数进行最优化求解，进而得到所有目标定标后的星等$m$。
在最优化求解过程中，这些参数会由于各种原因存在简并，如平场零点会同$a$项简并，大气质量变化较小时会导致$a$项和$kx$项简并。因此在实际定标时使用了一些先验知识和迭代的方法以破除参数间的简并（见表\ref{Fig:ubersdss}）。
下式中$m$是定标后的星等，$m_{ADU}$是仪器星等，下角标$\alpha$、$\beta$和$\gamma$分别代表$a$、$k$和$f$三项。
\begin{equation}
    m = m_{ADU} + a_{\alpha} - [k_{\beta} + (\frac{dk}{dt})_{\beta}(t-t_{ref})]x + f_{\gamma}(j)
    \label{Equ:Uber2}
\end{equation}


\begin{table}[ht!]
\footnotesize
\tabcolsep 31pt 
\begin{tabular*}{\textwidth}{cccc}
\toprule
    Parameter & Number & Fit & Comments \\ \hline
    a-terms & $6 \times 5 \times n_{night}$ & Yes & \\
    $k$ &  $5 \times n_{night}$ & Yes & k-term at $t = t_{ref}$\\
    $dk/dt$ & 5 & No & \\
    Flats fieldss & $6 \times 5 \times n_{season}$ & Yes(iterative) & 2048 element vector\\
    Amp-jumps & $6 \times 5 \times n_{run}$ & No & \\
\bottomrule
\end{tabular*}
\cnentablecaption{{\small SDSS Ubercalibration 定标模型中的参数\cite{Padmanabhan(2008)}。其中$n_{night}$表示每夜一次；$n_{season}$表示每个平场季一次；$n_{run}$表示每次扫描一次；$6 \times 5$表示5个滤光和6列相机片。$Amp-jumps$项表征了不同探测器之间相对增益的差异。© AAS. Reproduced with permission.}}{The parameters that make up the photometric model\cite{Padmanabhan(2008)}. 
The number of parameters is a function of $n_{night}$ (the number of nights), $n_{season}$ (the number of flat-field seasons), $n_{run}$ (the number of runs), and the number of filters (5) and camera columns (6).
The $Amp-jumps$ term is the relative gain of the two amplifiers. © AAS. Reproduced with permission.}
\label{Fig:ubersdss}
\end{table}

最终，通过Ubercalibration方法，SDSS巡天在$griz$波段的定标精度达到了1 \%左右，在$u$波段为2 \%左右。剩余误差主要是由Ubercalibration方法未考虑的大气消光短时标变化所导致的。
值得注意的是，由于SDSS的5个波段观测是准同时的，大气消光变化是偏灰的（随波长变化较小），
其短时标变化引入的误差在不同波段之间能较好地相互抵消，
所以SDSS颜色定标精度要优于其流量定标精度。

\subsubsection{Ubercalibration方法在PS1巡天中的应用} 

PS1巡天利用夏威夷1.8米望远镜面向3/4天区（Dec>$-$30度）使用五个滤光片（$grizy$）进行了大量重复观测\cite{Tonry et al.(2012)}。
Schlafly等人\cite{Schlafly(2012)}借助大量重叠观测数据，使用Ubercalibration方法对PS1进行流量定标。
该过程执行相对定标，鉴于PS1每个滤光片在相同区域反复观测约12次，且邻近区域有足够多的重叠观测源，因此整个观测区域的定标零点可以同时确定。

PS1的定标模型包含四项：$a$、$k$、$f$和$w$，其中$a$、$k$和$f$三项分别代表了光学系统和探测器的响应、地球大气消光以及平场。与SDSS巡天类似，$a$在每晚内不变，平场$f$在每一平场季（约一年）不变。
与SDSS巡天不同，大气消光系数$k$由于台站条件优秀在每晚内也假设不变。恒星点扩散函数的半高全宽会随成像质量发生改变。PS1巡天在测光时，过小和过大的半高全宽会造成流量的丢失，进而影响测光星等。因此$w$项代表像质造成的星等变化，用一个固定的二阶函数来描述，其常数项会同$a$项完全简并。因此，除了$w$常数项之外，所有参数均可以通过重复观测的结果进行拟合并加以限制。$a$和$k$也会在大气质量变化较小时发生简并，因此在定标中会使用先验值对$k$进行限制。
最终基于上述定标模型，$gri$波段的定标精度优于10 mmag，$zy$波段约为10 mmag，是定标精度最高的
地面巡天之一。

然而，该方法得到的PS1定标零点误差的空间分布呈现出大尺度和小尺度结构（如图\ref{Fig:delmag_radec}所示）。一个更为精确的定标过程将在4.6节介绍。

\subsubsection{Ubercalibration方法在XSTPS-GAC巡天中的应用} 
盱眙施密特望远镜反银心方向多色测光巡天利用紫金山天文台1.04/1.2米施密特望远镜对反银心方向约7000平方度天区开展了
g/r/i三个波段多色测光观测\cite{Liu et al.(2014)}，
为LAMOST反银心方向光谱巡天提供了高质量输入星表\cite{Yuan et al.(2015)LSS-GAC}。
为满足定标需求，相邻天区在赤经方向有50\%重叠，在赤纬方向有少许重叠，另外增加了Z字型天区（见苑海波等人（2015）图1\cite{Yuan et al.(2015)LSS-GAC}），连接相邻的赤纬天区，以增加在赤纬方向的重叠。在大量重复天区的帮助下，
并借助与SDSS天区的少量重叠，
我们利用Ubercalibration方法确定了每个天区的独立的零点参数。
重叠天区定标结果的一致性表明，XSTPS-GAC巡天定标精度与SDSS巡天相当。

\subsubsection{Ubercalibration方法在BASS巡天中的应用} 
北京-亚利桑那巡天（BASS）\cite{Zou et al.(2017)}利用基特峰2.3米Bok望远镜，对北银冠约5400平方度的区域进行了双色（g/r）测光观测。
为DESI项目\cite{DESI Collaboration et al.(2016)}选源提供了重要测光数据。
BASS巡天的大部分天区存在3次曝光，且每次曝光之间存在1/4视场的偏移。这种观测策略使得使得BASS可以有效地应用Ubercalibration方法开展内部定标，并利用PS1数据进行外部定标。通过这些重叠区域，BASS巡天对每
次曝光的零点进行了精确测量，并对每个观测季的平场改正误差利用恒星平场进行了修正。
最终，经过多次迭代，BASS巡天在$gr$两个波段对大部分天区实现了优于10mmag的定标精度\cite{ZhouZhimin(2018)}。

\subsubsection{Ubercalibration方法在{\it Gaia}巡天中的应用} 
由于欧空局{\it Gaia}卫星\cite{Gaia(2016)}是天基项目，因此在定标时不受地球大气消光所造成的影响，
其流量定标仅需考虑了仪器的影响。但是，由于其系统的复杂性（2个望远镜，106块CCD，G波段成像与BP/RP无缝光谱）
与探测器工作模式（16 种读出模式/曝光时长，3种读出窗口大小）的多样性，其定标模型的复杂性与参数的数目
\cite{Carrasco(2016)}是其它巡天所不能比拟的。
尽管如此，{\it Gaia}巡天是目前为止定标空间均匀性最高的巡天，其主要原因有三：
一是{\it Gaia}为天基项目，因此在定标时不受地球大气消光所造成的影响；
二是{\it Gaia}对全天不断重复扫描观测模式使得其天生特别适合于 Ubercalibration方法，可以用海量恒星的多次（最终约70次左右）重复观测对海量定标参数进行高精度限制，并进行定标结果的合并平均；
三是探测器采用时延积分模式（Time Delay Integration，TDI），因此，与SDSS巡天相同，{\it Gaia}的平场改正为一维向量，这降低了平场改正的难度。

在{\it Gaia} DR1中，对单次曝光该方法达到了3-4mmag的定标精度\cite{Evans(2017)}。在最新的{\it Gaia} EDR3中，其BP、RP、及G波段单次曝定标精度分别达到达到了3.0，1.8和2.4mmag\cite{Riello et al.(2021)}。
经数十次曝光叠加后，定标精度进一步提升到1mmag左右甚至更好，
与近期的一些独立检验结果一致 \cite{DESDR2(2021)}\cite{Niu et al.(2021c)}。


\subsubsection{Ubercalibration方法的局限性} 
\begin{enumerate}
  \item Ubercalibration方法限制定标参数时依赖于重复观测，其定标精度会受限于重复观测次数、
  重叠天区大小及重复观测策略\cite{Holmes et al.(2012)}，
  其定标精度也会同空间位置相关。比如SDSS巡天南北天区定标零点存在系统差异。
    \item 在对大气消光的参数化处理过程中，Ubercalibration方法忽略了一些可能有重要影响的效应，比如大气消光的短时标变化及大气消光的颜色项等。

    \item Ubercalibration方法假设所有定标星的能谱分布是相同的，这对于更高精度的定标来说不是一个好的假设。如果观测通带发生变化，需要进一步考虑其颜色改正。
  
    \item Ubercalibration方法是相对定标。
\end{enumerate}

综上，我们可知Ubercalibration方法忽略了一些次要的效应，需要在未来的定标工作中考虑。同样地，由于Ubercalibration该方法没有进一步精细地考虑大气变化所带来的影响，因此其地面定标精度止步于1\%左右。而为了突破地面测光巡天定标1\%精度的瓶颈，对大气消光更精细的处理必不可少。在后文中的Forward Global Calibration方法会对这一部分进行更加精细的建模，并得到了预期的效果。

\subsection{Hypercalibration方法} 
由于在不同巡天项目之间，大气条件、设备、观测和定标方式均有不同，因此不同巡天定标误差之间有相当强的独立性。
在上述假设下，Hypercalibration方法可以使用不同的巡天数据经颜色转换后进行相互校准\cite{Finkbeiner(2016)}。

\subsubsection{Hypercalibration方法在SDSS巡天中的应用} 
\begin{figure}[ht!]
   \centering
   \includegraphics[width=12cm]{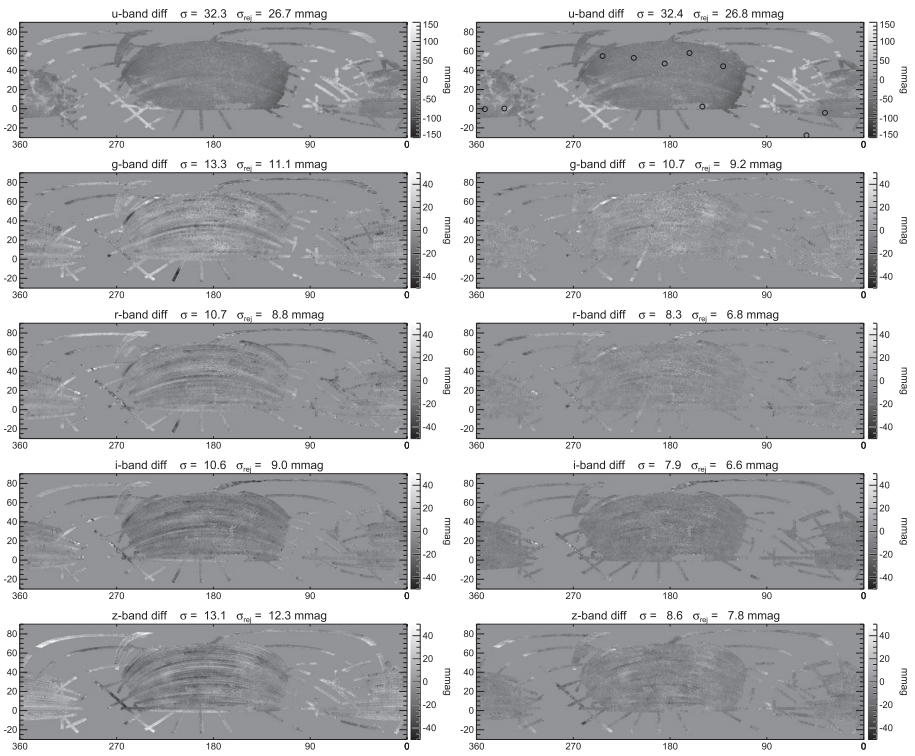}
   \cnenfigcaption{{\small PS1转换成的SDSS星等后和SDSS对应星等的比较结果\cite{Finkbeiner(2016)}。© AAS. Reproduced with permission.}} {{\small The mean difference of PS1 and SDSS, using the color transformations to convert PS1 to SDSS magnitudes. © AAS. Reproduced with permission.}}
  \label{Fig:Hyper1}
\end{figure}

由于SDSS的g/r/i/z测光系统和PS1的g/r/i/z测光系统比较接近，因此可以比较简单地在不同测光系统之间通过引入颜色项
进行转换。将基于PS1转换成的SDSS星等和SDSS的对应星等进行比较，可以观察到与SDSS观测模式相关的图样（如图\ref{Fig:Hyper1}所示）。
根据这些差异可以对SDSS数据的零点及平场进行修正。
最终，Hypercalibration方法将SDSS项目g/r/i/z波段的定标精度提升至9mmag，在高密度区域可以达到7mmag。 由于从PS1的g星等转换到SDSS的u星等是一种外推，受到金属丰度以及星际尘埃消光的强烈影响，因此无法用来修正SDSS u波段数据的零点。但考虑到金属丰度及消光变化的特征尺度远大于平场的特征尺度，
Hypercalibration方法可以对u波段的平场进行改正，精度约为15mmag。

\subsubsection{Hypercalibration方法的局限性与发展} 

Hypercalibration方法涉及到星等在不同测光系统之间的转换，其与恒星本身的物理性质有关，
只有在测光系统比较接近的时候才能保证比较高的精度。在测光系统偏差比较大时，星等转换本身
只考虑颜色项的情况下由于消光或星族性质的变化会带来比较严重的系统误差，
这也是为什么没有用PS1对SDSS u波段数据进行进一步零点修正的原因所在。
不过，在小天区范围内，在恒星消光及星族性质变化可以忽略的情况下， 只考虑颜色项的星等转化将不会带来明显的系统偏差。
此时，Hypercalibration方法可以用来改正相应尺度内定标零点的空间变化，如平场改正\cite{Thanjavur et al.(2021)}。

\subsection{Forward Global Calibration方法 } 
Stubbs等人2007年提出可以通过直接测量大气吸收谱的方式实现毫星等级别的定标精度\cite{Stubbs et al.(2007)}。这种方式需要确定瑞利散射、分子吸收、气溶胶和云这四个因素对大气吸收的贡献。从这个思路出发，FGCM方法通过对仪器（望远镜、滤光片、探测器）效应、大气（气体分子、臭氧、水蒸气、气溶胶）效应随时间和空间的变化进行参数化后进行正向建模，并结合外部辅助数据（仪器和大气监测数据）和足够多的重复观测源，继而通过最优化求解方法得到这些参数不同时刻的最佳值，最终对观测天区的数据进行定标。该方法\cite{Burke(2018)}由Burke等人提出，并应用于DES巡天。

FGCM方法定标后得到的每个波段$b$上的标准星等$m_b^{std}$，其定义同公式\ref{Equ:Standardsystem3}类似，即
\begin{equation}
    m_b^{std} = -2.5log_{10} (\frac{\int_{0}^{\infty} F_{\nu}(\lambda) \times S^{std}(\lambda) \times \lambda ^{-1} d \lambda}{\int_{0}^{\infty} F_{\nu}^{AB} \times S^{std}(\lambda) \times \lambda ^{-1} d \lambda})
\end{equation}
其中，$F_{\nu}^{AB} = 3631 Jy$（$1 Jy = 10^{−23} \, erg \, s^{-1} \, cm^{-2} \, Hz^{-1}$）\cite{Oke and Gunn(1983)}。
但$m_b^{std}$的通带是有明确定义的“标准通带”$S^{std}(\lambda)$，而实际观测星等$m_b^{obs}$其通带$S^{obs}(\lambda)$是随观测条件变化的。标准星等$m_b^{std}$和实际观测星等$m_b^{obs}$之差$\delta_b^{std}$可以通过FGCM方法建模得到。
这本质上是为了修正大气消光以及其它可能因素所造成的通带变化而进行的“颜色改正”（Chromatic correction）。 精确的颜色改正需要对天体在观测通带内的流量$F_{\nu}(\lambda)$和实际观测时的通带$S^{obs}(\lambda)$有细致的了解。
由于$F_{\nu}(\lambda)$无法从测光得到，因此FGCM方法使用了$F_{\nu}(\lambda)$的一阶线性展开，其常数项和一阶系数可以通过测光观测得到。而$S^{obs}(\lambda)$可以通过辅助设备的参数建模拟合得到。
如果$S^{obs}(\lambda)$和$S^{std}(\lambda)$一致或是$F_{\nu}(\lambda)$在观测通带的波长范围内不随波长变化时，该颜色修正项为零。

FGCM方法是对Ubercalibration方法的进一步发展，同时在定标过程中考虑了观测中测光系统响应曲线变化
（响应曲线随时间的变化、同种滤光片间的差异、滤光片不同位置响应差异等）对不同SED恒星的影响，
并能同时利用不同波段的观测数据。

\subsubsection{FGCM在DES巡天中的应用} 
DES项目利用托洛洛山美洲天文台(CTIO)4米布兰科望远镜的DECam相机对南银冠约5000平方度天区在$grizY$波段开展光学-近红外重复测光观测，极限星等约24等，以研究暗能量和暗物质的性质。这里以DES项目为例，介绍FGCM方法的建模过程。

FGCM方法在应用于DES项目当中时，首先对于$F_{\nu}(\lambda)$项进行了简化处理。由于天体在观测通带波长范围内的具体流量分布无法直接从宽带测光中得到，因此FGCM方法使用$F_{\nu}(\lambda)$的一阶展开代替$F_{\nu}(\lambda)$。每个观测通带内$F_{\nu}(\lambda)$的一阶展开系数将由相邻通带的标准星等和波长计算得到。因此，观测通带的波长范围越窄，目标天体在其内的流量分布变化越小，简化的准确度越高。

对于实际观测时的通带$S^{obs}(\lambda)$，FGCM方法将$S^{obs}(\lambda)$分为了仪器和大气这两个相互独立的部分，
如下式所示：
\begin{equation}
    \begin{split}
    S_b^{obs}(\lambda) &= S_b(x,y,alt,az,t,\lambda) \\
    &= S_b^{inst}(x,y,t,\lambda) \times S^{atm}(alt,az,t,\lambda)，
    \end{split}
\end{equation}
式中x和y是目标在焦平面上的位置；alt和az是望远镜指向的高度角和方位角。
将仪器效应和大气效应分离后，前者如式\ref{Equ:FGCM2}所示：
\begin{equation}
    \begin{split}
    S_b^{inst}(x,y,t,\lambda) &= S_b^{flat}(pixel,epoch) \times S_b^{starflat}(pixel,epoch) \\
    &\times S_b^{superstar}(ccd,epoch) \times S^{optics}(MJD)  \times S_b^{DECal}(ccd,\lambda)， \\
    \end{split}
    \label{Equ:FGCM2}
\end{equation}
其中，仪器响应考虑了：flat、star-flats、Superstar Flats、optics、DECal五部分。
flat为平场；star-flats为恒星平场，通过多次拍摄密集星场得到，用来矫正平场的大尺度结构畸变；Superstar Flats进一步对star-flats进行优化，反映了不同CCD之间的灵敏度差异；optics用来估计光学系统由于环境尘埃在裸露光学仪器表面上的积累以及主镜镀层退化导致的效率下降，与波长无关；DECal代表通过单色照明系统测量得到的包括主镜反射率、滤光片通带和探测器效率的系统效率随波长的变化。
对于这五部分，其变化的时标较长，参数通常在数个月至一年才会有明显的变化，而且除optics外均为测量量。
因此，唯一需要拟合的参数是optics。由于主镜镜面会进行周期性清洗，FGCM方法考虑了其相对于每次清洗时
尘埃积累和主镜镀层变化导致的光学系统效率的线性下降，因此对于每个清洗周期optics有两个拟合参数。
所以，对于仪器响应总体而言，FGCM方法每个清洗周期只有这两个拟合参数。

由于大气消光因包含多个成分，
精确改正比较复杂。FGCM方法需要考虑的因素较多，拟合参数也相应变多，如式\ref{Equ:FGCM3}所示：
\begin{equation}
    \begin{split}
    S_b^{atm}(alt,az,t,\lambda) &= S_b^{molecular}(bp,zd,t,\lambda)  \times S^{pwv}(zd,t,\lambda) \times e^{-X(zd)\times \tau(t,\lambda)}
    \end{split}
    \label{Equ:FGCM3}
\end{equation}
其中，bp是大气压，zd是天顶距。
FGCM方法考虑了三部分的影响：molecular、pwv和$e^{-X\tau}$。
其中，molecular代表空气中除水汽和气溶胶之外的分子瑞利散射和吸收效应，除了臭氧的吸收每晚上需要一个自由参数来描述之外，其余可以通过辅助设备对大气压的实时测量得到。


pwv代表水汽的影响，其数据可以从辅助设备中得到，但并不是每次曝光都有辅助设备数据。
当辅助设备数据可用时，pwv表达式由公式\ref{Equ:FGCM4}给出，其中拟合参数$pwv_0$是每晚的零点，$pwv_1$是一阶项的常数。
当辅助设备数据不可用时，pwv表达式由公式\ref{Equ:FGCM5}给出，每晚随时间线性变化，拟合参数$pwv$和$pwv_s$将由时间UT进行拟合。
 \begin{equation}
     \begin{split}
     PWV(exposure) &= pwv_0(night) + pwv_1(run)  \times pwv_{aux}(exposure)
     \end{split}
     \label{Equ:FGCM4}
 \end{equation}
 \begin{equation}
     \begin{split}
     PWV(exposure) &= pwv(night) + pwv_s(night) \times UT(exposure)
     \end{split}
     \label{Equ:FGCM5}
 \end{equation}

气溶胶组分非常复杂，为简单起见，在当前的FGCM方法中，气溶胶散射简化为单一微粒的气溶胶，
用两个参数$\tau _{7750}$和$\alpha$来描述，如下式所示
\begin{equation}
    \tau(\lambda) = \tau _{7750} \times (\lambda / 7750 \mathring{A})^{-\alpha} 
    \label{Equ:FGCM6}
\end{equation}
其中$\tau _{7750}$为归一化系数，每晚内随时间变化。$\alpha$与气溶胶颗粒尺寸和形状有关，
不同晚上取值不同，但每晚内不变。对于$\tau_{7750}$变化的处理方式同水汽类似，
有无辅助设备数据的情况有$\tau _0(night)$、$\tau _1(run)$、$\tau(night)$和$\tau _s(night)$四个拟合参数，加上$\alpha$本身有5个拟合参数。

综上，对大气消光每晚上需要8个自由参数来描述，如公式\ref{Equ:FGCM7}所示。注意其中$pwv_1(run)$和$\tau_1(run)$不随
每晚上变化。给定这些参数之后，可以利用地球大气模型（如MODTRAN\cite{Berk et al.(1999)}）计算得到地球大气的消光曲线。
 \begin{equation}
     \emph{P}^{atm} == (O_3, pwv_0, pwv_1, pwv, pwv_s, \tau _0, \tau _1, \tau, \tau _s,\alpha)
     \label{Equ:FGCM7}
 \end{equation}


FGCM方法在定标开始时，对于待定标的星、曝光及测光夜有三条筛选标准：待定标星在$griz$每个波段至少要有两次观测；待定标的曝光要有至少600颗待定标星；在此基础上需要待定标的测光夜至少有10次曝光。在满足这些条件后，下一步将会利用筛选得到的定标星迭代拟合出最佳的仪器参数${P}^{inst}$和大气参数${P}^{atm}$及每颗星每次曝光的$m_b^{std}$。在迭代结束后，
更新Superstar Flats和每次曝光的灰消光值。
其中灰消光值指的是每次曝光中定标星残差的平均值，通过灰消光值的分布可以剔除如有云或者显著仪器错误的曝光图像。
将这些有问题的曝光剔除后，FGCM会进行下一轮循环。在DES的Y3A1（前三年）数据中，FGCM方法循环了4次。通过同一颗星每次拍摄时$m_b^{std}$的弥散，可以估计出FGCM方法的内部精度，其在DES的Y3A1数据中可以稳定地达到6-7mmag的内部一致性。

由于FGCM方法依赖于重复观测，因此在最新的DES DR2 \cite{DESDR2(2021)}中，基于DES项目前六年的观测数据的DR2内部一致性经平均后有进一步提升。
在把$g/r/i$星等合并转换为$G$星等并同{\it Gaia} DR2 \cite{Gaia(2018)} $G$星等比较，在15角分的分辨本领下其随空间位置的弥散为2.15mmag。
假设这三个波段相互独立，那么DES DR2单个波段内部定标的精度好于$2.15 \times \sqrt{3}$=3.6mmag。

\subsubsection{FGCM方法的局限性与发展} 
FGCM方法相对于Ubercalibration方法，对地球大气及仪器响应随时间、空间及波长变化进行了更精细的处理，在有许多
简化的情况下也取得了预期的效果，但是在其实际应用时面临较高的门槛：
\begin{enumerate}
    \item 同Ubercalibration方法类似，要求观测时存在大量时间和空间相重合的观测。
    \item 依赖于辅助设备对大气进行高精度监测，依赖于单色光源对仪器响应曲线的测量。
    \item 需要测光夜观测，一个晚上需要足够数量的曝光次数，需要多个波段观测。
\end{enumerate}

此外，FGCM方法并不能进行绝对定标，绝对定标仍需要结合外部的标准星来进行。对于DES来说，其使用的是HST\cite{Bohlin(2007)}的标准星。
FGCM方法在未来将应用于Vera C. Rubin Observatory Legacy Survey of Space and Time\cite{LSST(2019)}(LSST)项目中。

\section{软件驱动的方法}
“软件驱动”类方法则是基于对天体性质，特别是恒星颜色的认识，包括
Stellar Locus Regression (SLR\cite{High et al.(2009)}）、
Stellar Locus (SL\cite{Lopez-Sanjuan et al.(2019)}）以及 
Stellar Color Regression (SCR\cite{Yuan et al.(2015a)})方法。

\subsection{SLR方法}
\begin{figure}[ht!]
   \centering
   \includegraphics[width=8cm]{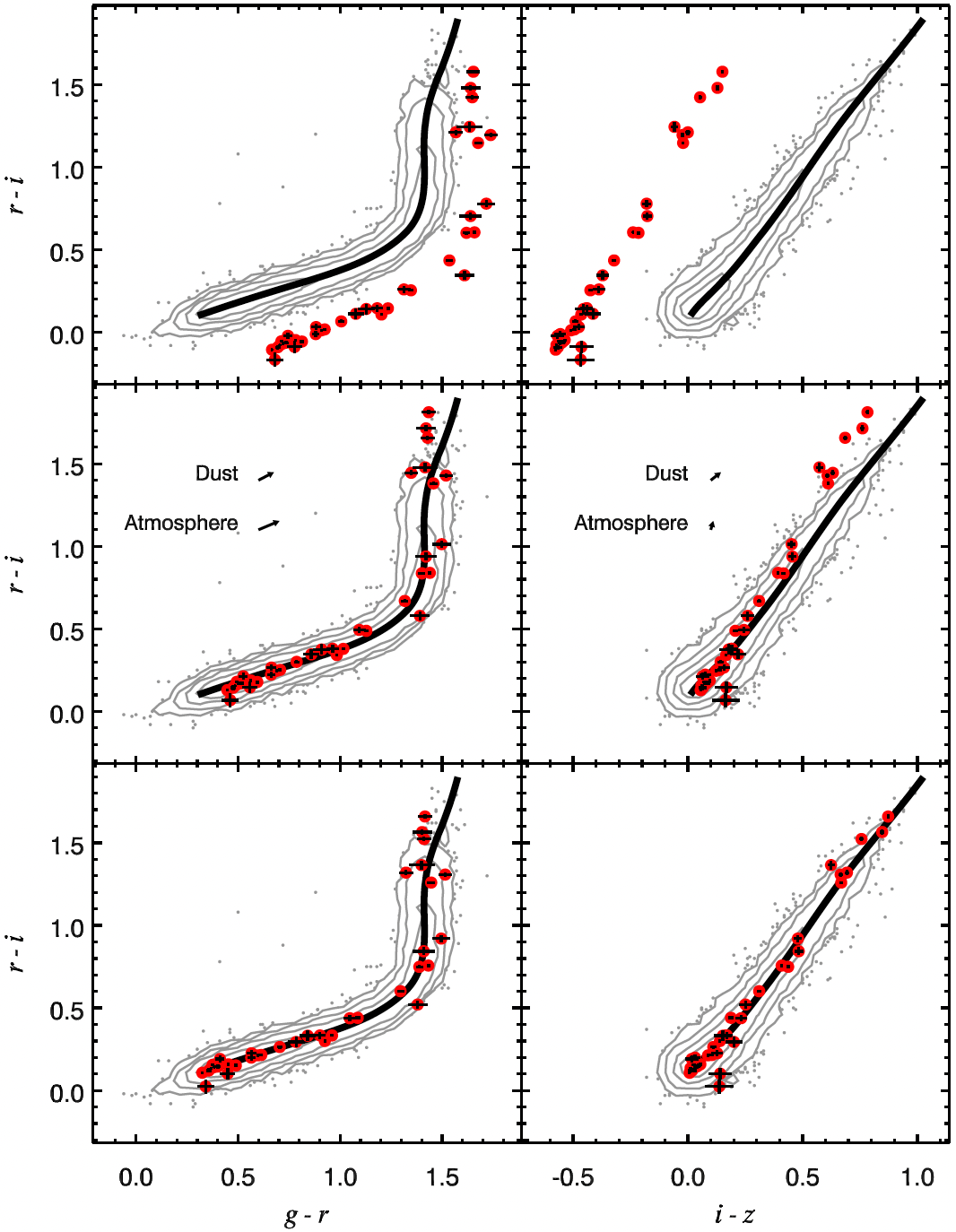}
   \cnenfigcaption{{\small SLR方法示意图\cite{High et al.(2009)}。每幅子图中黑线和灰色密度等值线代表标准恒星轨迹，红点为IMACS恒星的颜色分布。顶部子图展示的是定标前状态（IMACS仪器颜色）；中间子图展示的是将颜色空间中的恒星分布与标准恒星轨迹相匹配后的结果（忽略颜色项），矢量显示了星际尘埃(Ar=0.2)和大气(1.3大气质量)消光的预期方向和幅度；底部子图展示的是考虑颜色项改正后的匹配结果。© AAS. Reproduced with permission.}} {{\small Illustration of the SLR method\cite{High et al.(2009)}. All panels show the standard stellar locus by black lines and gray density contours. And the red dots are stellar colors obtained from IMACS. The top panels show the stellar without calibration(the instrumental IMACS colors); the middle panels show the result of SLR, which is performed with only a common translation vector applied to the instrumental colors
   (ignoring the color term), with vectors showing the expected direction and magnitude of the extinction of interstellar dust (Ar = 0.2) and atmosphere (1.3 air masses); the bottom panels show the result with the correction of the color term.  © AAS. Reproduced with permission.}}
  \label{Fig:slr}
\end{figure}

通过研究恒星在内禀颜色-颜色空间的分布性质，发现大多数恒星都位于一维恒星颜色轨迹（Stellar locus)上，其位置主要取决于恒星的有效温度。
Ivezi{\'c}等人\cite{Ivezic et al.(2004)}首次利用恒星颜色轨迹的位置来估计SDSS巡天的测光零点。
Covey等人\cite{Covey et al.(2007)}通过SDSS和2MASS的约600,000个高质量恒星样本，确定了$ugrizJHK_{s}$随
$g-i$变化的恒星颜色轨迹。
SLR方法假设恒星的颜色轨迹是普适的，通过将仪器颜色空间中的恒星分布与标准恒星轨迹相匹配，使得我们能够用一组简化的步骤取代传统的分析过程。基于Covey等人给出的恒星颜色轨迹，该方法应用到麦哲伦 6.5米望远镜IMACS设备拍摄的部分天区数据上，其精度可达百分之几\cite{High et al.(2009)}。
Gilbank等人\cite{Gilbank et al.(2011)}采用一个非常类似的方法应用到The second Red-sequence Cluster Survey (RCS-2)中，
在$griz$波段取得了优于3\%的颜色定标精度。
Bleem等人\cite{Bleem et al.(2015)}将该方法应用到Blanco Cosmology Survey\cite{Desai et al.(2012)}中，在$griz$波段取得了优于2\%的颜色定标精度。
图\ref{Fig:slr}展示了SLR方法的实现过程。该方法可以可靠定标恒星和星系的颜色，实时改正仪器零点、大气不透明度变化及银河系星际红化带来的影响。

SLR方法的局限性体现在以下几个方面：
\begin{enumerate}
    \item 至少需要三个波段观测数据，且包含至少一个蓝端的滤光片（如$g$波段）来打破定标过程中的简并。
    \item 只适用于消光均匀的低消光区域。
    \item 没有考虑金属丰度、表面重力加速度的影响（恒星颜色轨迹不是严格普适的）。
    \item 相对于星等定标，SLR方法只能给出相对精确的颜色定标。
\end{enumerate}

\subsection{SL方法}
SL方法\cite{Lopez-Sanjuan et al.(2019)}
是对SLR方法的进一步发展。该方法结合已定标好的数据和消光信息，利用恒星颜色轨迹估计出恒星在任意波段的星等，将其作为流量标准星来定标。
SL方法已应用到J-PLUS巡天的定标过程中：结合PS1 DR1\cite{Tonry et al.(2012)}、{\it Gaia} DR2\cite{Gaia(2018)}测光数据和3D消光图\cite{Green et al.(2018)}得到J-PLUS DR1\cite{Cenarro et al.(2019)}的各波段星等（比如PS1的$(g-i)_0$内禀颜色和$r$星等得到J-PLUS DR1的$J0660$星等）。在金属丰度敏感的波段（$u$、$J0378$、$J0395$），该方法取得的定标精度约为2\%，其他波段（$g$、$J0515$、$r$、$J0660$、$i$、$J0861$、$z$）定标精度约为1\% \cite{Lopez-Sanjuan et al.(2019)}。
该方法打破了SLR方法中至少需要三个波段观测且只能做颜色定标的限制，然而不同滤光片之间恒星颜色转换受金属丰度的影响仍未考虑。

L{\'o}pez-Sanjuan等人借助于PS1 DR1、{\it Gaia} DR2测光数据及LAMOST DR5提供的金属丰度信息，利用SL方法对J-PLUS DR2进行了定标\cite{Lopez-Sanjuan et al.(2021)}。该过程讨论了金属丰度对利用SL方法对JPLUS DR2数据定标的影响，在改正金属丰度导致的零点误差后，J-PLUS DR2的定标精度达到0.01mag左右。

\subsection{SCR方法} 
近年来，随着多目标光纤光谱巡天的快速发展，如SDSS/SEGUE\cite{Yanny et al.(2009)}, LAMOST\cite{Deng et al.(2012)}\cite{Liu et al.(2014)}\cite{Zhao et al.(2012)}, APOGEE\cite{Jonsson et al.(2020)}， and GALAH\cite{De Silva et al.(2015)}，我们进入了百万乃至千万量级恒星光谱的时代。
基于现代模板匹配以及数据驱动的恒星参数测量软件\cite{Lee et al.(2008a)} \cite{Lee et al.(2008b)} \cite{Wu et al.(2011)} \cite{Xiang et al.(2015)} \cite{Xiang et al.(2017)},
我们可以得到非常高内部精度的恒星大气参数({\teff}, {\logg}及{\feh})\cite{Niu et al.(2021a)}。因此，恒星的 内禀颜色可以基于大范围光谱巡天数据，并结合恒星配对方法\cite{Yuan et al.(2013)}精确预言。SCR方法即以数百万经过光谱观测的星作为颜色标准星，在进行高精度的消光改正后预测恒星观测颜色，并同实测颜色进行比较以改正地球大气及仪器效应所带来的各种影响。

SCR方法的基本原理可以通过下式表示
\begin{equation}
     Color_{\rm obs}  = Color_{\rm intr} + Reddening
\end{equation}
上式中，$Color_{\rm intr}$为恒星的内禀颜色，$Color_{\rm obs}$为恒星的观测颜色，$Reddening$为该颜色星际红化值。

SCR方法的基础是单颗恒星的内禀颜色是相对比较简单的，可以由恒星的少量几个内禀性质决定，
如有效温度、金属丰度、表面重力加速度等。
SCR方法的核心是利用基于已有数据对恒星内禀性质的理解来限制恒星的内禀颜色。
SCR方法可以有多种形式（黄博闻 \& 苑海波 2022)\cite{Huang and Yuan(2021)}: 
\begin{enumerate}
    \item 恒星内禀颜色来自大规模光谱巡天得到的恒星大气参数。这一过程可以是基于恒星配对方法\cite{Yuan et al.(2013)}，也可以是基于简单的多项式拟合关系（肖凯 \& 苑海波 2022）\cite{Xiao and Yuan(2022)}，或者是一些基于机器学习的回归算法。对于宽波段颜色来讲，通常仅需考虑有效温度、金属丰度、表面重力加速度的影响。对于一些窄波段颜色来讲（如JPLUS巡天中的J0515波段），还需考虑单个元素丰度的影响。
    \item 恒星内禀颜色来自大规模光谱巡天得到的恒星光谱。基于机器学习方法可以实现从恒星观测量（光谱）到另一恒星观测量（内禀颜色）的预测。这种形式的好处是跟恒星理论大气模型无关。
    \item 恒星颜色来自恒星多波段颜色。所用恒星多波段颜色（如$U-B, B-V, V-R, R-I$)包含有恒星基本性质的信息，特别是金属丰度, 所以可以用来精确预测恒星其它波段的颜色\cite{Yang et al.(2021)}。
    \item 恒星内禀颜色来自丰度依赖的恒星颜色轨迹。恒星丰度信息可以来自大规模光谱巡天\cite{Huang et al.(2021a)}，也可以来自大规模测光巡天。
\end{enumerate}

SCR方法的另一核心问题为红化改正，涉及我们对银河系尘埃分布（消光值）与性质（消光系数）的理解。
消光值的来源主要包括基于恒星配对方法得到的消光与银河系二维消光图\cite{Schlegel et al.(1998)} \cite{Planck Collaboration et al.(2014)} \cite{Irfan et al.(2019)}。在特殊情况下也可用银河系三维消光图\cite{Green et al.(2018)}\cite{Chen et al.(2019)}来代替，但精度有限。
消光系数通常可以通过恒星配对方法经验得到，也可假设一定的消光规律通过理论计算得到。

随着{\it Gaia} DR2 和 EDR3\cite{Gaia(2016)} \cite{Gaia(2018)} \cite{Gaia(2021)}数据的释放，我们可以获取到均匀而精确的高质量全天测光数据，其在$G$、$BP$和$RP$波段的精度达到了前所未有的毫星等级别。
通过在颜色中引入{\it Gaia}的测光数据，SCR方法可以精确预言恒星在不同波段的星等，使得颜色标准星成为流量标准星，并开展高精度流量定标。

\section{恒星颜色回归方法的应用、问题与发展} 
在大视场巡天时代，SCR方法借助大规模光谱巡天提供的海量精确恒星大气参数和{\it Gaia}
高精度测光数据，构建了精度在1\%左右的千万量级的“标准星”。 其海量高精度标准星为在流量定标过程中改正地球大气的短时标变化，
改正望远镜、滤光片及探测器的各种大尺度效应，打破地面测光巡天的1\%精度瓶颈奠定了基础。
SCR方法已经在多个巡天中进行了应用，并对原先定标的系统误差成功进行了证认、溯源与改正，
大幅提升了巡天的定标精度。
因此，本节将介绍SCR方法在实际中的应用情况，并对下一步的发展方向进行讨论。

\subsection{Stripe 82颜色定标} 
SDSS Stripe 82标准星表\cite{Ivezic et al.(2007)}是SDSS巡天定标最好的星表，从实际上定义了SDSS的测光系统。
苑海波等人\cite{Yuan et al.(2015a)}首次提出SCR方法，并结合SDSS的恒星大气参数应用于该标准星表数据中，
对大气消光改正误差及平场改正误差进行了精确测量，
成功将$u-g$、$g-r$、$r-i$和$i-z$四种颜色的定标精度分别提升至5、3、2、2mmag，是之前精度的2--3倍。
我们发现该星表不同颜色平场改正误差之间存在非常强的相关性与线性关系，
其根源来自于原先平场改正过程中主成分颜色
（principle colors，为2个颜色的线性组合得到）的应用，
其表现与预期很好地吻合（详见论文\cite{Yuan et al.(2015a)}3.6节)。
研究同时发现，在SDSS巡天中，其$z$波段的部分CCD中存在非线性效应。

基于重新定标后的标准星表数据，我们研究了主序恒星在SDSS测光系统中恒星颜色轨迹的丰度依赖性，
经验上证明了其内禀宽度在考虑丰度影响后几乎为零\cite{Yuan et al.(2015b)}；
发展了基于对恒星颜色轨迹的偏差无偏测量大样本恒星双星比例的方法，证实了双星比例与金属丰度的反相关，发现银晕中包含更高比例的双星系统\cite{Yuan et al.(2015c)};
发展了基于丰度依赖的恒星颜色轨迹测量恒星丰度的方法，精度与低分辨光谱巡天相当\cite{Yuan et al.(2015d)}；
研究了巨星的恒星颜色轨迹的丰度依赖性及基于丰度依赖的恒星颜色轨迹区分巨星并测量丰度的方法\cite{Zhang et al.(2021)}。 这些工作一方面说明了我们对定标误差估计的可靠性，另一方面也表明了高精度测光数据在
恒星性质测量方法的重要作用。

\subsection{{\it Gaia} DR2 and EDR3颜色修正} 
{\it Gaia} DR2 和 EDR3\cite{Gaia(2016)} \cite{Gaia(2018)} \cite{Gaia(2021)}提供了精确且均匀的全天测光数据，
在$G$、$G_{\rm BP}$和$G_{\rm RP}$波段其测光精度达到了前所未有的毫星等量级。
为了覆盖从6等到22等较宽的星等范围，{\it Gaia}依据$G$星等亮度不同使用了不同的观测模式，
对应不同的积分时间与读出窗口大小。 不同观测模式之间的流量定标是相对独立的，
这导致其星等和颜色存在随亮度变化的系统误差
\cite{Maiz Apellaniz and Weiler(2018)}\cite{Casagrande and VandenBerg(2018)} 
，并严重限制了{\it Gaia}测光数据本应有的潜力。

\begin{figure*}[ht!]
   \centering
   \includegraphics[width=15.cm]{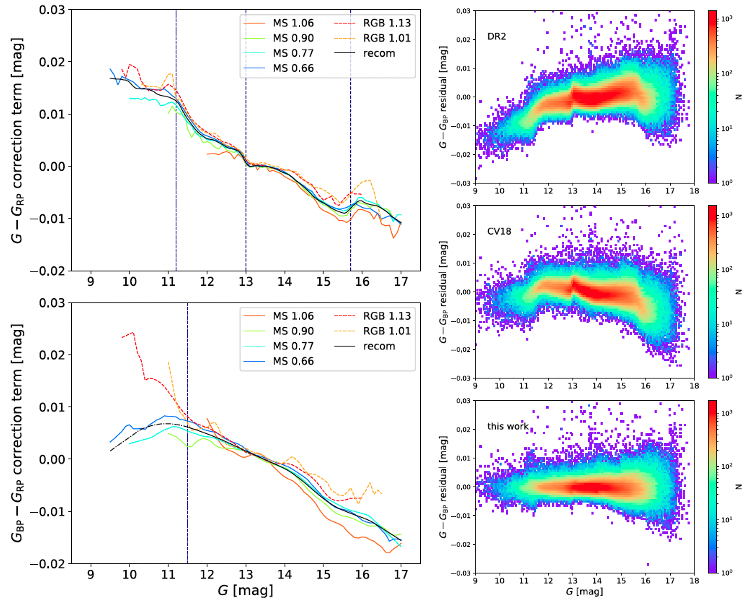}
   \cnenfigcaption{{\small 左侧两幅图分别展示了{\it Gaia} DR2中$G-G_{\rm RP}$和$G_{\rm BP}-G_{\rm RP}$颜色定标误差随$G$星等的变化趋势；
   右侧三幅图展示了$G-G_{\rm RP}$颜色残差随$G$星等的分布，上图是DR2发布的数据、中图是经CV18\cite{Casagrande and VandenBerg(2018)}修正后的结果、下图是SCR方法修正后的结果\cite{Niu et al.(2021a)}。© AAS. Reproduced with permission.}} {{\small The left panels show the color-dependent and recommended calibration curves for $G-G_{\rm RP}$ and $G_{\rm BP}-G_{\rm RP}$ with $G$ magnitude in {\it Gaia} DR2, respectively; the right panels are the 2D histogram of $G-G_{\rm RP}$ residual after subtracting the metallicity-dependent color locus with $G$ magnitude. Top panel: published DR2 data. Middle panel:  the same after applying CV18\cite{Casagrande and VandenBerg(2018)} corrections. Bottom panel: applying corrections from the SCR method\cite{Niu et al.(2021a)}. © AAS. Reproduced with permission.}}
  \label{Fig:gaiadr2}
\end{figure*}

牛泽茜等人\cite{Niu et al.(2021a)}\cite{Niu et al.(2021b)}利用LAMOST DR5和DR7提供的高质量大气参数，将SCR方法分别应用到{\it Gaia} DR2(如图\ref{Fig:gaiadr2}所示) 和EDR3巡天数据中，对{\it Gaia}颜色的定标误差随星等及颜色的变化进行了精细的刻画。
牛泽茜等人从LAMOST-{\it Gaia}共同源中选取观测质量足够高且消光值足够小的数据，将其分为主序矮星和红巨星样本，
选择其中$G$星等在13.3到13.7范围内的恒星作为参考样本，利用该参考样本建立恒星大气参数（{\teff}, {\logg}及{\feh}）与内禀颜色之间的关系。接着将该关系应用于全体样本并改正尘埃消光以后，得到了观测颜色和模型颜色之差随G星等的变化趋势，最后将这一趋势作为修正曲线。
结果表明，{\it Gaia} DR2数据中不同观测模式间转换会带来较为明显的颜色跳变，整体最大差别可达20毫星等。{\it Gaia} EDR3数据很大程度上减小了各种定标误差，但整体趋势仍然可见。经过修正后的颜色定标精度可以达到1毫星等，为进一步利用{\it Gaia}测光数据研究个体恒星和群体样本的统计性质奠定了基础\cite{Niu et al.(2021c)}\cite{Xu et al.(2022)}。

\subsection{{\it Gaia} EDR3星等修正}
\begin{figure*}[ht!]
   \centering
   \includegraphics[width=15.cm]{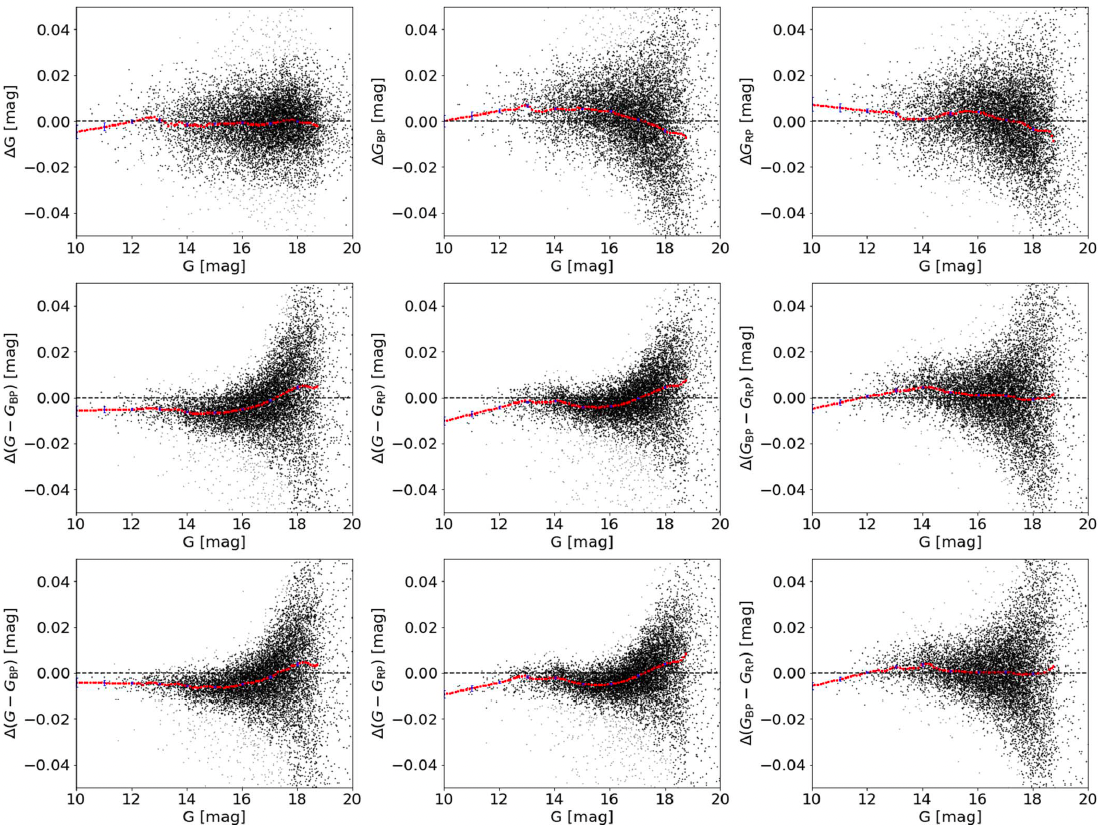}
   \cnenfigcaption{{\small 颜色位于$G_{\rm BP}-G_{\rm RP}\in [0.6, 1.5]$的源的残差随$G$星等的分布\cite{Yang et al.(2021)}。
   顶部三幅图是星等残差的分布，中间三幅图展示的颜色由上面三幅图得到，底部三幅图展示的颜色由神经网络直接训练得到。每一个面板中，$G$星等在12.5-18.8 mag范围内分格，宽度为0.4 mag步长为0.1 mag；通过3$\sigma$原则剔掉灰色偏离点后得到每个格中的中值。在亮端，
   对$G<13.0 \, mag$的星进行线性拟合，并把线性拟合的结果外推
   于$G<12.5 \, mag$的情况，然后采用LOWESS对校准曲线做平滑处理（frac = 0.07）。结果用红色虚线表示，并基于500个子样本通过Bootstrap方法估计出误差条（蓝色）。© AAS. Reproduced with permission.}} {{\small Residual distributions as a function of G magnitude for $G_{\rm BP}-G_{\rm RP}\in [0.6, 1.5]$ \cite{Yang et al.(2021)}.
   The top panels show the distribution of the residuals with magnitude, the middle three panels show result of the colors calculated from the top panels, and the bottom three panels show the colors directly trained from the neural networks. 
   For each panel, stars are divided into different bins of width of 0.4 mag at a step size of 0.1 mag when $12.5 < G < 18.7$ mag. The median value for each bin is estimated, with a 3 − $sigma$ clipping performed and the gray dots dropped. At the bright end, a linear fitting is performed for stars of $G$ < 13.0 mag, the results are adopted when $G$ < 12.5 mag. Finally, a locally estimated scatterplot smoothing (LOWESS) is applied to smooth the calibration curves, with frac = 0.07. The final results are indicated by red dotted lines. The blue error bars are estimated with 500 subsamples using the Bootstrap method. © AAS. Reproduced with permission.}}
  \label{Fig:GaiaEDR3}
\end{figure*}

杨琳等人\cite{Yang et al.(2021)}
利用仔细挑选的10000颗左右具有精确$UBVRI$星等测量的Landolt标准星\cite{Clem & Landolt(2013)}，
通过一种机器学习的方法精确估计这些标准星在{\it Gaia}三个波段的精确星等。
在估计的过程中，仔细考虑了恒星金属丰度和消光对恒星颜色的影响。通过这种方式，我们精确测量了{\it Gaia} EDR3星等定标系统误差在$10<G<19$等区间内随星等的修正曲线(如图\ref{Fig:GaiaEDR3}所示)。
与4.2节结果相比，我们基于完全不同的数据和方法得到的结果一致性好于1个毫星等。在修正曲线的基础上，我们还利用哈勃空间望远镜得到的CALSPEC光谱库\cite{Bohlin et al.(2014)}获取了EDR3波段的绝对零点。
该结果为充分利用{\it Gaia}精确星等与距离信息奠定了基础，为以{\it Gaia}星等为基础定标其它巡天数据扫清了障碍。

\subsection{Stripe 82流量定标} 

随着{\it Gaia} DR2 和 EDR3\cite{Gaia(2016)} \cite{Gaia(2018)} \cite{Gaia(2021)}数据的释放，我们可以获取到均匀而精确的高质量全天测光数据，其在$G$、$G_{\rm BP}$和$G_{\rm RP}$波段的精度达到了前所未有的毫星等级别。
通过结合{\it Gaia}的测光数据，将待定标波段和Gaia的测光数据组合成颜色，
SCR方法可以精确预言恒星在不同波段的星等，这使得高精度流量定标成为可能。

\begin{figure*}[ht!]
   \centering
   \includegraphics[width=15.cm]{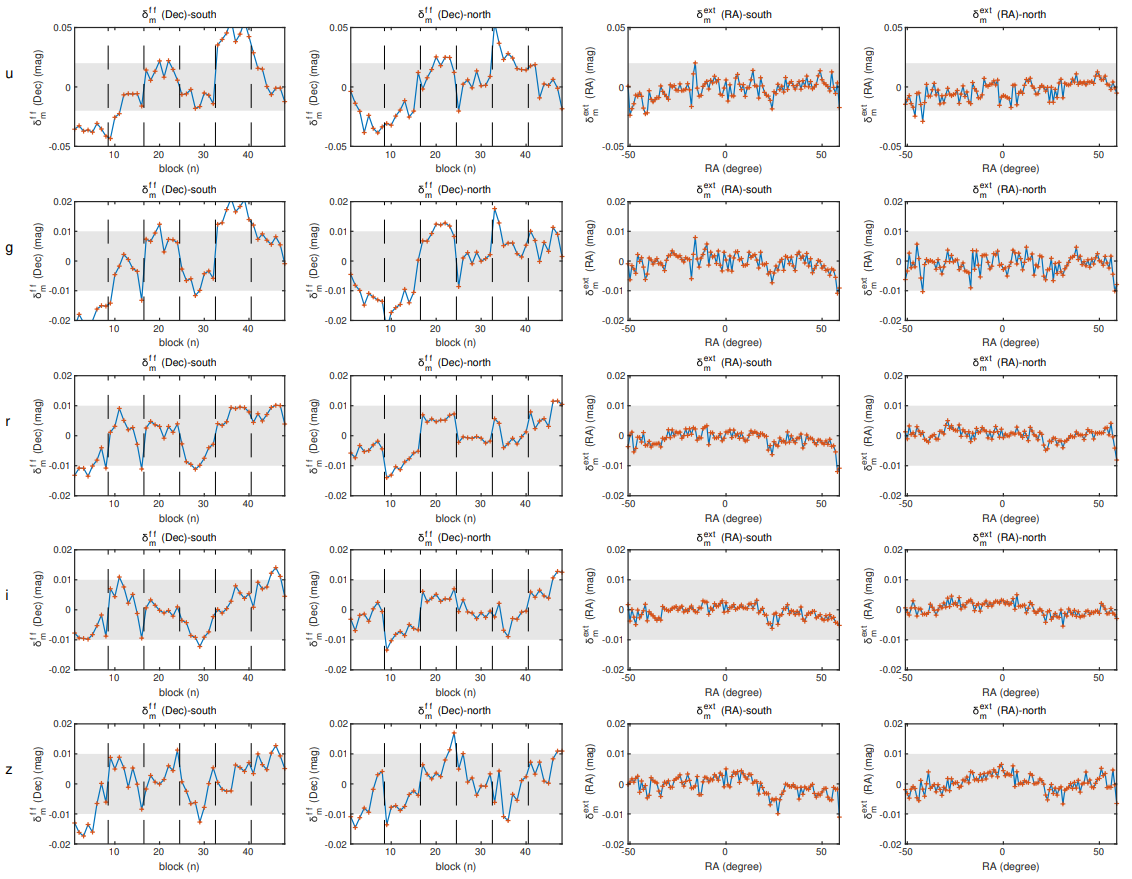}
   \cnenfigcaption{{\small SCR方法给出的SDSS Stripe 82天区V2.7版本标准星表在赤经和赤纬上的独立修正结果\cite{Huang and Yuan(2021)}。其中$\delta _m^{ext}(R.A.)$和$\delta_m^{ff}(Dec.)$便分别是$ugriz$五个波段赤经和赤纬上的独立修正值。参考图\ref{Fig:SDSS}中SDSS巡天的观测模式，为了填补CCD之间的空隙，SDSS巡天在SDSS Stripe 82天区沿赤经方向扫描观测时会进行南北两次扫描。图中第1列和第2列分别为南侧和北侧条带中$\delta_m^{ff}(Dec.)$在赤纬方向上的变化，第3列和第4列为分别为南侧和北侧条带中$\delta_m^{ext}(R.A.)$在赤经方向上的变化。
   另外，图中第1列和第2列的垂直虚线表示了不同CCD之间的大致边界。灰色阴影区域在$u$波段表示$\pm$0.02mag，在$griz$波段表示$\pm$0.01mag。© AAS. Reproduced with permission.}} {{\small Independent corrections to the SDSS Stripe 82 region of V2.7 standard catalog at R.A. and Decl. from the SCR method\cite{Huang and Yuan(2021)}. Where $\delta _m^{ext}(R.A.)$ and $\delta_m^{ff}(Dec.)$ are the independent corrections in the R.A. and Decl. directions for the five bands of $ugriz$, respectively. 
   Referring to the observation of the SDSS survey in Fig. \ref{Fig:SDSS}, in order to fill the gap between CCDs, the SDSS survey will perform two scans(north and south) when drift scanning along the R.A. in the SDSS Stripe 82 region. 
   The first and second columns plot the $\delta_m^{ff}(Dec.)$ as a function of Decl. for south and north strips, respectively. The third and fourth columns plot the $\delta_m^{ext}(R.A.)$ as a function of R.A. for south and north strips, respectively.
   In addition, the vertical dashed lines mark the approximate boundaries between the different camera CCD columns. The shaded region are of $\pm$0.02mag in the $u$ band and $\pm$0.01mag in the $griz$ bands. © AAS. Reproduced with permission.}}
  \label{Fig:stripe82}
\end{figure*}

结合上小节修正后的{\it Gaia} EDR3星等以及LAMOST DR7和SDSS DR12的光谱数据，黄博闻和苑海波将SCR方法应用于SDSS Stripe 82 V2.7版本标准星表\cite{Ivezic et al.(2007)}并对其进行了重新流量定标\cite{Huang and Yuan(2021)}。
我们提供了Stripe 82天区在赤经方向$1^{\circ}$、赤纬方向上约$0.03^{\circ}$分辨本领下的星等修正，
分别对应原先大气消光改正与平场改正误差，结果如图\ref{Fig:stripe82}所示。
通过比较LAMOST DR7和SDSS DR12两种不同恒星大气参数来源所得到的结果的一致性、Stripe 82南北strip相同CCD平场改正间的一致性及不同波段平场改正间的一致性，发现在$u$波段定标精度达到5mmag，在$griz$波段达到约2mmag，是之前结果的5倍。

我们还将SCR方法应用于最新的V4.2星表\cite{Thanjavur et al.(2021)}中。
最新的V4.2星表基于{\it Gaia} EDR3的测光数据，使用了类似于Hypercalibration的方法将{\it Gaia}的测光系统转化至SDSS测光系统对Stripe 82进行定标。我们发现其在赤经方向仍然存在由于消光和星族的变化导致的较为明显的定标零点偏差。在赤纬方向，其对小尺度平场做出了较好的修正，与预期一致。

\subsection{SMSS DR2} 
\begin{figure*}[ht!]
   \centering
   \includegraphics[width=15.cm]{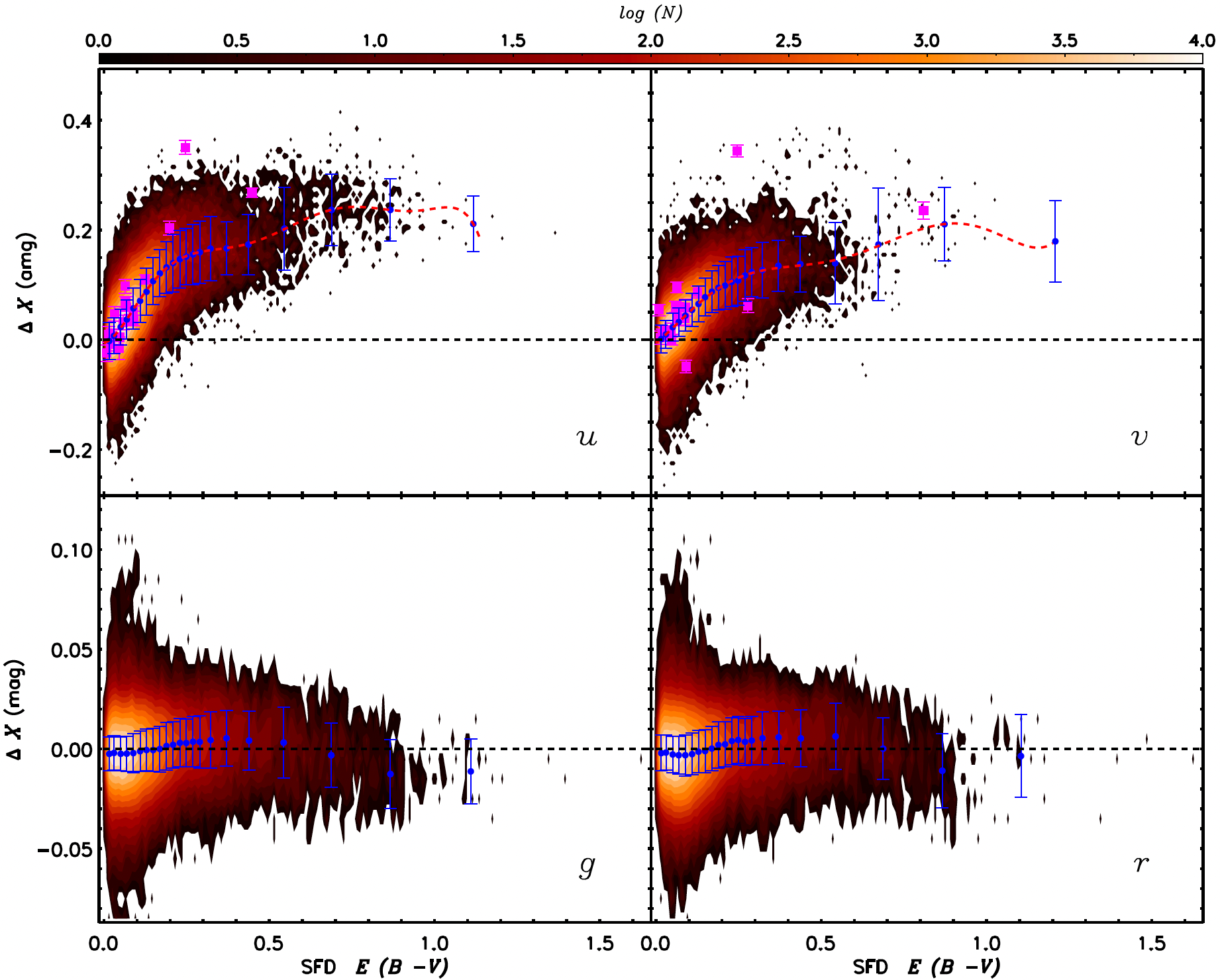}
   \cnenfigcaption{{\small SMSS DR2在$uvgr$波段的星等零点误差随$E(B-V)_{\rm SFD}$的分布\cite{Huang et al.(2021a)}。在$E(B-V)_{\rm SFD}$上分格后，格中的均值在图中用蓝点表示；图中蓝色误差条代表格中数据的标准偏差。上面两幅图中的红色虚线是蓝点的最佳拟合曲线；粉红色方块代表由NGSL库中恒星计算出的AB星等与SMSS观测值之差随$E(B-V)_{\rm SFD}$的分布。描述数据点个数的颜色条在顶部给出。© AAS. Reproduced with permission.}} {{\small Magnitude offsets, as a function of $E(B-V)_{\rm SFD}$, for $uvgr$ in SMSS DR2\cite{Huang et al.(2021a)}. The blue dots in each panel represent the mean values of each $E(B-V)_{\rm SFD}$ bin; the error bars indicate the standard deviations of the magnitude differences of each $E(B-V)_{\rm SFD}$ bin. The dashed red lines in the top two panels represent the best fits to the median differences as a function of $E(B-V)_{\rm SFD}$;
   the magenta squares represent the difference between the synthetic AB magnitudes computed for stars in the NGSL library and the SkyMapper observed values, as a function of $E(B-V)_{\rm SFD}$. The color bar describing the number of data points is shown at the top. © AAS. Reproduced with permission.}}
  \label{Fig:smss}
\end{figure*}

SkyMapper Southern Survey\cite{Wolf et al.(2018)}(SMSS)
是澳大利亚主导的一个针对河内的南天测光巡天项目，其
最大特色在于配备了对恒星参数非常敏感的$u,v$滤光片。
黄样等人\cite{Huang et al.(2021a)}基于{\it Gaia} DR2测光数据和GALAH DR3\cite{Buder et al.(2021)}提供的恒星大气参数，
使用SCR方法对SMSS DR2\cite{Onken et al.(2019)}进行了重定标。
其$uvgr$波段预测星等通过丰度依赖的恒星颜色轨迹方式获得，在此过程中消光与消光系数均来自恒星配对方法。
研究发现，在SMSS DR2的原始定标过程中存在零点误差，其在对重力和金属度敏感的紫外波段尤为突出。
如图\ref{Fig:smss}所示，零点误差与消光存在强相关，在$E(B-V)\sim 0.5$时，零点误差分别达到0.174($u$波段)和0.134 mag ($v$波段)。
该零点误差来源于SMSS DR2根据ATLAS参考星表\cite{Tonry et al.(2018)}进行流量定标时的尘埃项。而对于$gr$波段，
由于星等转换过程中尘埃项的系数很小，零点误差随$E(B - V)$的变化可以忽略不计。
经过改正与尘埃消光相关的零点偏差之后，剩余零点偏差在$uvgr$波段仍显示出了一定幅度的空间变化特征。
由于定标星数量有限，我们对空间变化特征也进行了简单的多项式修正。最终取得的定标精度
$uv$波段在1\%左右，$gr$波段在0.5\%左右。

\subsection{PS1 DR1} 
\begin{figure*}[ht!]
   \centering
   \includegraphics[width=15.cm]{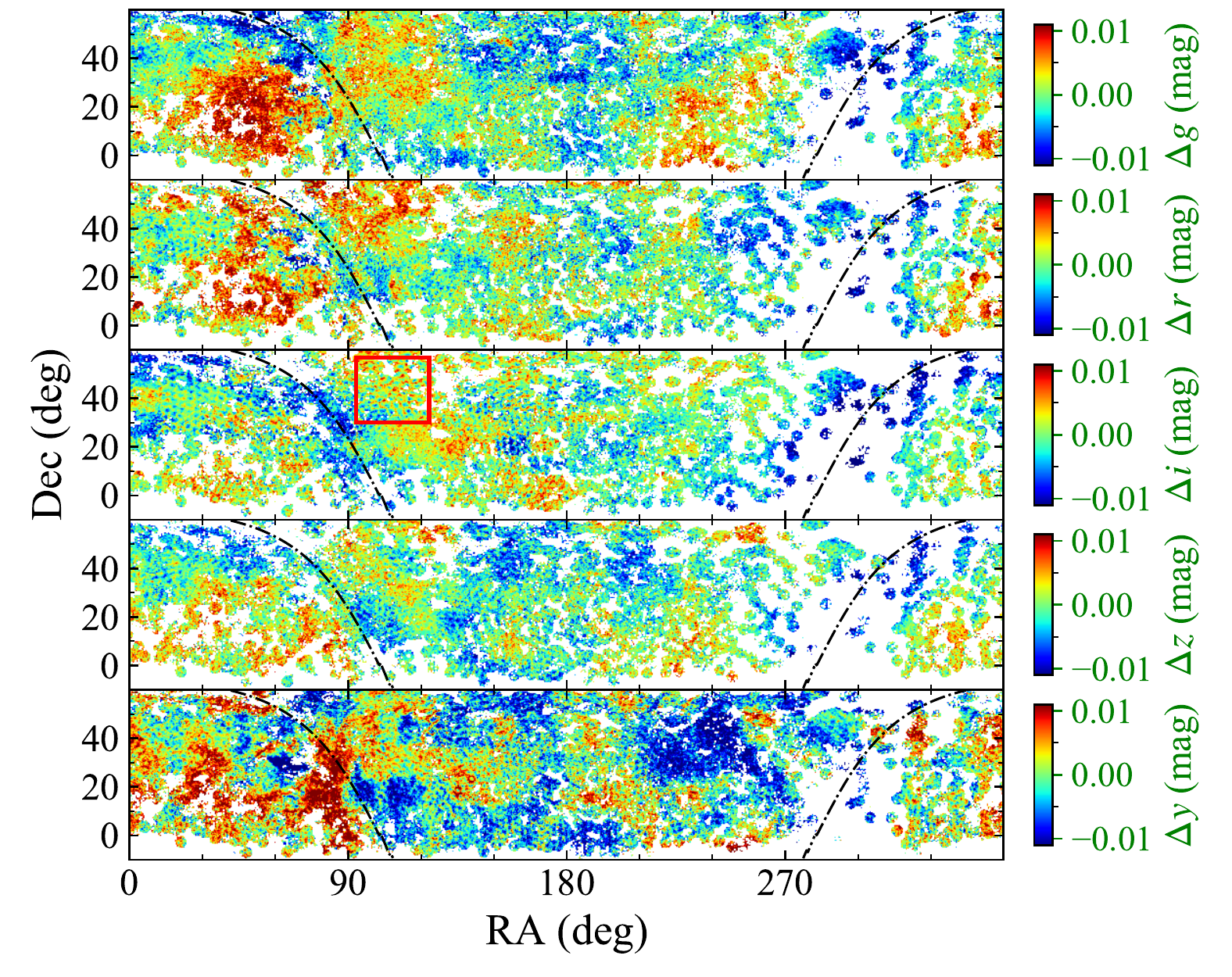}
   \cnenfigcaption{{\small 经过$20' \times 20'$合并之后的模型星等与观测星等之差在(R.A., Dec.)上的空间分布\cite{Xiao and Yuan(2022)}。从上到下分别对应于$g$, $r$, $i$, $z$ 和 $y$五个波段。每个子图中的黑色虚线代表银盘，右侧颜色棒表征着差值大小。© AAS. Reproduced with permission.}} {{\small Spatial variations of magnitude offsets(model magnitudes and observed magnitudes) after a $20'\times 20'$ binning in the $g$, $r$, $i$, $z$ and $y$ bands\cite{Xiao and Yuan(2022)}. The black dotted line indicates the Galactic plane in each panel. Color bars are overplotted to the right. © AAS. Reproduced with permission.}}
  \label{Fig:delmag_radec}
\end{figure*}

借助于修正后的{\it Gaia} EDR3精准测光数据和LAMOST DR7光谱数据，
肖凯和苑海波使用SCR方法对PS1 DR1\cite{Tonry et al.(2012)}的$grizy$波段的流量定标进行了独立检验和重定标\cite{Xiao and Yuan(2022)}。
研究发现，在 $grizy$ 波段PS1流量定标精度在每20角分的分辨本领下分别为4.6、3.9、3.5、3.6和4.8 mmag。
另外，如图\ref{Fig:delmag_radec}所示，SCR方法预言的模型星等与观测星等之差在五个波段皆呈现出由PS1定标误差导致的大尺度和小尺度的空间变化（高达1\%）。
我们还发现PS1 DR1星等存在与星等相关的系统误差，在 $grizy$ 波段分别为5、5、5、4和3 mmag/mag，这可能来源于点扩散函数（PSF）测光误差。
我们提供了非均匀空间分辨率（从20$'$到160$'$）的二维分布图，用于修正星等差的空间变化。

\subsection{SAGE DR1}
Stellar Abundance and Galactic Evolution\cite{Zheng et al.(2018)}\cite{Zheng et al.(2019)}\footnote{Fan Zhou et al. in preparation}（SAGE）巡天采用一套对恒星大气参数敏感的独特测光系统，包括$u_s$、$v_s$、$g$、$r$、$i$、$H\alpha_{\rm n}$、$H\alpha_{\rm w}$、$DDO51$八个滤光片。SAGE巡天利用美国2.3米Bok望远镜、新疆天文台南山1米望远镜、
乌兹别克斯坦1米望远镜及紫金山天文台盱眙1.04/1.2米施密特望远镜，对北天约1.2万平方度高银纬天区展开巡天观测。
其中，最重要的$u_s$和$v_s$两个波段观测由Bok望远镜完成，其相机采用4块CCD拼接而成，每块探测器有4个读出通道。

对$u_s$和$v_s$波段，SAGE DR1的定标过程对每个通道分别进行，定标策略如下：
1）在LAMOST DR5、{\it Gaia} DR2、SAGE DR1共同源的基础上，
选择标准星（$5300<T_{\rm eff}<6800$ K、$-0.8<{\rm [Fe/H]}<0.2$和snr>20），
并构建SAGE星等与LAMOST大气参数和{\it Gaia}星等之间的关系，例如$BP-u_s=f(T_{\rm eff}, {\rm [Fe/H]})$；
2）得到每幅图、每个通道上的零点；
3）在每个通道上用关于在CCD上位置坐标的二阶多项式进行平场改正，继而通过迭代得到新的零点和平场；
4）重新估计每个通道上的零点；
5）结合Ubercalibration方法通过重叠观测源对不包含或包含少量LAMOST标准星的图像进行定标；
6）重新构建SAGE星等与LAMOST提供的大气参数和{\it Gaia}星等之间的关系，并带回第2）步进行迭代；
7）通过经验光谱库合成得到的与观测的颜色-颜色关系（例如$BP-u_s~vs.~BP-RP$）的对比实现绝对定标。
上述过程应用于SAGE DR1 $u_s$/$v_s$波段，定标精度在5 mmag左右\footnote{Yuan Haibo et al. in preparation}。

\subsection{JPLUS DR1 and DR2}
Javalambre Photometric Local Universe Survey\cite{Cenarro et al.(2019)}(JPLUS)
利用西班牙83厘米巡天望远镜计划对北天8500平方度天区使用12个光学滤光片
（$uJAVA$、$J0378$、$J0395$、$J0410$、$J0430$、$g$、$J0515$、$r$、$J0660$、$i$、$J0861$和$z$）展开巡天观测。
其中，7个窄带滤光片针对恒星的重要光谱特征，对恒星参数非常敏感。
JPLUS DR1\cite{Cenarro et al.(2019)}包含1000多万颗源的观测数据，巡天面积覆盖约1022平方度。 
JPLUS DR2 \footnote{Varela \& J-PLUS collaboration, in preparation}覆盖约2200平方度天区。

L{\'o}pez-Sanjuan等人借助PS1 DR1\cite{Tonry et al.(2012)}测光数据，利用SL方法对JPLUS DR1数据进行了原始定标，定标精度约为10-20 mmag\cite{Lopez-Sanjuan et al.(2019)}。
苑海波等人\footnote{Yuan Haibo et al. in preparation}借助{\it Gaia} DR2测光数据和LAMOST DR5光谱数据，使用SCR方法对JPLUS DR1数据进行了重定标。研究发现：
JPLUS DR1原始定标过程中存在随空间变化的零点误差，且在不同波段呈现出很强地相关性。
该误差在$uJAVA$、$J0378$、$J0395$、$J0410$和$J0430$波段约为10-27 mmag，在其余波段约为4 mmag。
误差主要来源于如下三个方面：
1）金属丰度的影响，尤其是在金属丰度敏感的波段（蓝端）；
2）3D消光图\cite{Green et al.(2018)}的误差；
3）红化系数的误差。
经零点及平场误差改正后，最终每个波段取得的定标精度约为2-5 mmag。
苑海波等人基于重新定标后的JPLUS DR1数据，借助于{\it Gaia} DR2测光数据和LAMOST DR5提供的光谱数据，发展了利用机器学习估计恒星基本参数的方法\cite{Yang et al.(2021b)}，并通过该方法获得了约200万颗恒星的有效温度（$\delta T_{\rm eff}\sim 55$ K）、表面重力加速度（$\delta \log g\sim0.15$ dex）、金属丰度（$\delta {\rm [Fe/H]}\sim0.07$ dex）以及$\rm \alpha$、$\rm C$、$\rm N$、$\rm Mg$、$\rm Ca$元素丰度（$\delta \sim55$ dex）。这为银河系化学动力学分析提供了优秀的数据集。

L{\'o}pez-Sanjuan等人借助于PS1 DR1测光数据、{\it Gaia} DR2距离数据及LAMOST DR5光谱数据，利用考虑金属丰度影响后的SL方法对JPLUS DR2数据进行原始定标，
定标精度约为10 mmag\cite{Lopez-Sanjuan et al.(2021)}。
苑海波等人同样对JPLUS DR2数据通过SCR方法进行重定标。发现JPLUS DR2定标精度在考虑金属丰度影响有有显著提升，在$uJAVA$、$J0378$、$J0395$、$J0410$和$J0430$波段零点误差约为6-14 mmag，其余波段约为3 mmag。但是，仍然存在随空间变化的零点误差，可能来源于SFD消光图。

\subsection{恒星颜色回归方法的问题与发展} 

借助大规模高精度光谱巡天数据和{\it Gaia}高精度测光数据，SCR方法在改正每个时刻由于大气或仪器原因导致的零点变化及
大尺度平场改正方面展现出了强大的能力，能比较常规地突破地面测光巡天1\%精度的瓶颈，达到数个毫星等级别的内部定标精度。
为实现1个毫星等的定标精度，该方法仍有进一步的提升空间:
\begin{enumerate}
   
    \item 经典SCR方法依赖于光谱巡天数据。目前，使用SCR方法构建的具有精确恒星参数的标准星数目受限于光谱巡天的天区面积、巡天深度及样本大小。比如当前最主要的LAMOST巡天数据主要集中在北天，其暗端极限星等亮于17.8等。而新一代测光巡天比如LSST、CSST亮端饱和星等均在18等左右，且主要在南天或南北天兼顾。
    随着DESI\cite{DESI Collaboration et al.(2016)}、WEAVE\cite{Dalton et al.(2012)}、4MOST\cite{de Jong et al.(2016)}及LAMOST二期工程等新一代光谱巡天的开展，将会在光谱数量、巡天深度以及巡天面积等方面有进一步质的提升（如光谱数量增加到上亿级别，极限星等增加到21等左右，巡天面积则南北天兼顾），届时将为SCR方法的普遍应用并实现1个毫星等的定标精度提供强大的数据保障。与此相关，新一代测光巡天如能在策略上同时开展一个较浅的巡天，如连接起LAMOST巡天深度与其标准巡天深度，则将非常有助于其定标工作的开展\cite{Huang et al.(2021a)}。

   \item SCR方法所预言的星等是基于参考场中定义的“标准响应曲线”得到的星等，尚未考虑大气消光导致的颜色效应。在将来应用的过程中，特别是针对偏蓝的宽波段和受水汽吸收影响比较大的波段，在比较模型星等与观测星等时
   应考虑定标零点可能随大气不透明度、大气质量及天体颜色的变化。
   注意SCR方法当前所用定标星通常为FGK型恒星，所以得到的定标零点对于更蓝或更红的天体可能带有一定的偏差。
    
    \item SCR方法构建的标准星的精度受限于红化改正的精度。当前高精度消光改正还面临三个方面的挑战：
    1）消光规律的空间变化；
    2）相同消光规律下红化系数随消光值以及恒星参数（如温度）的变化
    （尤其在紫外波段和宽波段更为明显\cite{Niu et al.(2021a)}\cite{Niu et al.(2021b)}）；
    3）当前常用的二维或三维消光图本身随尘埃热发射参数及空间位置变化的系统偏差\cite{Sun et al.(2022)}
    幸运的是，
    LAMOST等大规模光谱巡天数据通过恒星配对等方法对数百万视向方向上恒星多波段消光的精确测量也为解决上述挑战提供了重要的机遇\cite{Sun et al.(2022)}\footnote{Zhang Ruoyi et al. in preparation}。

   \item 由于FGK型恒星在紫外波段颜色与恒星参数非常敏感，且受到恒星活动性的影响，所以基于SCR方法估计FGK恒星紫外波段颜色精度较差。为进一步提升空间紫外波段（如CSST NUV波段）的定标精度，SCR方法将利用白矮星和早型主序星作为定标源。
   首先，这些恒星紫外辐射强且稳定；
   其次，这些恒星紫外颜色受金属丰度的影响非常小，非常有利于高精度的颜色估计；
   再次，目前的巡天数据已发现大量的白矮星\cite{Gentile Fusillo et al.(2018)}和早型主序星，有足够的定标源的数量；
   最后，这些恒星的大气参数测量已取得较高的精度\cite{Xiang et al.(2021)}\cite{Sun(2021)LAMOSTpara}。
   
   \item 诸多例子已充分说明了SCR方法的能力。但是，目前实践大都是针对已定标数据的重新定标\footnote{这并不意味着SCR方法只能对已定标数据进行重定标。SCR方法与Ubercalibration等方法一样，可以从原始数据开始进行内部定标，比如4.7节介绍的SAGE巡天。}，许多观测信息非常难以利用。考虑到长时标变化量（如平场相关参数）可以利用更多的数据进行更精确的限制，在巡天初始定标的时候就引入SCR方法能更好的发挥该方法的优势，进一步提升最终定标的精度。

\end{enumerate}

随着{\it Gaia} DR3数据释放，我们将获得约一亿颗G星等亮于17.6等恒星的BP、RP无缝光谱数据。
CSST也计划获取数亿条恒星的从近紫外到近红外波段无缝光谱数据。
在好的定标情况下，无缝光谱数据包含精确且丰富的恒星参数信息\cite{Yang et al.(2021b)}。
基于无缝光谱得到的恒星参数或直接基于无缝光谱数据直接预测星等信息都将为SCR方法注入新的活力。

亮红星系的内禀颜色性质也相对比较简单，主要由红移决定，同时受到星系质量/光度的影响\cite{Peek and Graves(2010)}。当前大规模宇宙学巡天也已获得了海量亮红星系的光谱，基于亮红星系做为颜色标准星开展高精度颜色定标也值得进一步研究。

\section{总结与展望}
均匀且精确的流量定标是大视场测光巡天的难点和成功的关键因素之一。针对测光巡天流量定标的难点，
人们发展了多种方法，“硬件驱动/观测驱动”类以FGCM方法为代表，“软件驱动/物理驱动”类以SCR方法为代表，已成功突破1\%内部精度的瓶颈，并朝着更高的目标逐步迈进。
随着巡天数据的累积，特别是新一代测光（如LSST、CSST）和分光（如DESI、LAMOST二期工程）巡天的开展；随着对大气消光、仪器效应更好的理解（如\cite{Xiao2021}）；随着对银河系不同类型恒星性质更好的测定，大视场
巡天流量定标的精度将越来越高。
不同的定标方法之间有非常好的互补性，在测光巡天定标时，针对巡天的特点，
建立最合适的定标模型（如大气消光模型、平场改正模型、仪器响应曲线模型等），
有机结合各种方法提供的对模型的限制，必将推动测光巡天流量定标精度整体进入毫星等时代。


可以预期，在定标精度接近毫星等的情况下，PSF测光系统误差及红化改正误差对我们
研究天体的内禀性质，特别是亮源的内禀性质会产生更重要的影响，需要更多的关注。
高精度测光数据的科学应用也值得更深入挖掘,
如\cite{Niu et al.(2021c)}\cite{Xu et al.(2022)}\cite{Huang et al.(2022)}。


\Acknowledgements{
感谢云南大学刘晓为教授将作者引入天文实测领域及在
本文撰写过程中提供的大量宝贵意见与建议。
}

{}

\makeentitle

\authorcv[]{Bowen HUANG}{}

\authorcv[]{Kai XIAO}{}


\authorcv[]{Haibo YUAN}{}
\authorcv[]{}{}

\end{document}